C:/COMETS/ABCs/Draft10/130510

# "The location of Asteroidal Belt Comets (ABCs), in a comets' evolutionary diagram: The Lazarus Comets"

This Arxiv.org version of the manuscript, contains new material not in the printed version of the paper: a) a crust thickness calculation for two α values, and b) inclusion of comet C/2012 T1 Panstarrs, bringuing the total number of members of the ABC group to 12.

Ignacio Ferrín, Jorge Zuluaga, Pablo Cuartas,
Institute of Physics,
Faculty of Exact and Natural Sciences,
University of Antioquia,
Medellin, Colombia
ferrin@fisica.udea.edu.co

| | |
|---|---|
| Number of pages | 40 |
| Number of Figures | 13 |
| Number of Tables | 5 |

**Abstract**

There is a group of newly recognized asteroids in the main belt that are exhibiting cometary characteristics. We will call them Asteroidal Belt Comets or ABCs for short. The surprising property of these objects is that their orbits are entirely asteroidal while their behavior is entirely cometary, with Tisserand invariants larger than 3.0, while all Jupiter family comets have Tisserand invariants smaller than 3.0. An analysis of their orbital and physical properties has resulted in the following conclusion:

(1) We define the *"detached group, DG"* as those objects that exhibit cometary characteristics (sublimating water) and have aphelion distances $Q < 4.5$ AU. The detached group contains all the ABCs traditionally recognized, plus a few other members not traditionally recognized like 2P and 107P. With the above definition there are 12 members of the ABC group: 2P, 107P, 133P, 176P, 233P, 238P, C/2008 R1, C/2010 R2, 2011 CR42, 3200, 300163 = 2006 VW139 and P/2012 T1 Panstarrs. And there are three members of the collisioned asteroids, CA, P/2010 A2, 596 Scheila and P/2012 F5 Gibbs.

(2) In the literature a common reason for activity is interplanetary collisions. Active objects sublimate ices except for P/2010 A2, 596 Scheila, and P/2012 F5 that have exhibited dust tails due to collisions and 3200 Phaethon activated by solar wind sputtering. In this work we will trace the origin of activity to a diminution of their perihelion distances, a hypothesis that has not been previously explored in the literature. We propose that the asteroidal belt contains an enormous graveyard of ancient dormant and extinct rocky comets that turn on (are rejuvenated) in response to a diminution of the perihelion distance, caused by planetary perturbations.

(3)We have calibrated the black body (color) temperature of comets vs perihelion distance, R, regardless of class. We find $T = 325 \pm 5\ ^{o}K/\sqrt{R}$.

(4) Using a mathematical model of the thermal wave we calculate the thickness of the crust or dust layer on comet nuclei. We find a thickness of 2.0±0.5 m for comet 107P, 4.7±1.2 m for comet 133P and 1.9±0.5 m for a sample of 9 comets. Notice the small errors.

(5) We have located three ABCs in an evolutionary diagram of Remaining Revolutions, RR, vs Water-Budget Age, WB-AGE. ABCs lie together in the upper right hand corner of the diagram, as expected from physical arguments.

(6) The RR vs WB-AGE comparison also defines the region of the *graveyard of comets,* as those objects with 1000 cy < WB-AGE, where cy stands for *comet years*. Five members belong to the graveyard, 107P, 133P, 2006 VW139, D/1891 W1 Blanpain and 3200 Phaeton. Thus we propose that the asteroidal belt contains an enormous graveyard of ancient dormant and extinct rocky comets, that turn on (are rejuvenated), in response to a diminution of their perihelion distance.

## 1. Nomenclature

Comets in the asteroidal belt have been referred to in the literature with several names. Main belt comets, MBC, was used by Jewitt (2010a) at first to describe anything that showed activity. This was later used ambiguously, sometimes generally, for those objects that showed extended emission, other times for those shown to be periodically active and assumed to be driven by sublimation. Active main belt objects, AMBOs, used by Bauer et al. (2011) was meant to eliminate the ambiguity of the MBC term by defining anything that showed activity that was in the main belt. Activated asteroids, AA, coined by Licandro et al. (2009) refers to centaur comets (Active Centaurs, AC) and active NEAs (like 2009 WJ50). In this work we will use asteroidal belt comets or ABCs to describe a subset of the AMBOs, at least three of which have been identified as being active due to collisions. This defines a subgroup of collisioned asteroids, CA. However we believe the nomenclature "ABC" reflects more clearly the primordial nature of the active comets. In resume ABC is a subset of AMBO, which in turn is a subset of AA. If we restore the initial significance for MBC then MBC = ABC. The difference between AMBO and the DG is that the DG is a cuantitative definition, while AMBO is a general definition. Then AMBO = the Detached Group, DG, with $Q < 4.5$ AU . The following equations summarize the nomenclature issue:

$$\begin{aligned} DG &= AMBO = ABC + CA \\ AA &= DG + AC = ABC + CA + AC \end{aligned} \tag{0}$$

## 2. ABCs

(A) Introduction

In this work it will be shown that the activity of ABCs is due to a diminution of their perihelion distance, q, that produces an energy enhancement, EE, proportional to $1/q^2$ . This in turn means that the thermal wave penetrates deeper into the nucleus, sublimating layers of ice that were previously beyond the wave's reach. This process will be called *rejuvenation*. ABCs are located in a comet's evolutionary diagram of Revolutions Remaining vs Water Budget Age. They occupy the upper right hand corner of the diagram, and an explanation is given of why they reside there.

(B) Membership.

To define families of objects we will use the aphelion distance of the orbit, Q. For a comet to move into the inner solar system, a major planet has to perturb it into a new orbit whose new aphelion is the old perihelion. Thus Q classifies objects by parent perturbing planet.

The inclusion by Jewitt and Li (2010) of 3200 Phaeton into the ABC list is an interesting decision. In their paper on 3200 they conclude that Phaeton is essentially an extinct rocky comet, and their inclusion opened the door to include other members (see Figure 1).

The *detached group of comets* will be defined as those with aphelion distances Q<4.5 AU. That means that no planet perturbed them in, and thus they must be autochthonous. Their orbits are decoupled from Jupiter. The detached group contains all the ABCs traditionally recognized, plus a few other members not traditionally recognized like 2P and 107P. With the above definition there are 11 members of the detached group: 2P, 107P, 133P, 176P, 233P, 238P, C/2008 R1, C/2010 R2, 2011 CR42, 3200, and 300163 = 2006 VW139. While writing this paper a new member of the detached group was recognized: P/2012 F5 Gibbs, but it does no belong to the ABC group. It belong to the CA group.

Their physical and orbital properties are compiled in Table 1, and their location and that of Jupiter family comets is shown in the eccentricity, e, vs semi-major axis, a, diagram (Figure 1). This diagram also explains the selection of the cut-off distance Q<4.5 AU. It encompases all AMBOs and leaves out all Jupiter Family comets.

(C) Origin of activity.

Currently a preferred mechanism for activation is collision with an interplanetary body. Activation by collision has been demonstrated for P/2010 A2 (Jewitt et al. 2010a; Snodgrass et al., 2010) a comet that exhibited a detached striated tail with no corresponding coma due to water sublimation. Additionally P/2010 A2 has been shown to be an S-type asteroid (Kim et al., 2012), while comets are type C, or D. Other proposals for activation have been electrostatic levitation and rotational bursting (Jewitt, 2010b) but these have not been demonstrated. Since P/2010 A2 does not sublimate water it should not be considered an ABC and should be dropped off the list (Table 1) (Kim et al., 2012; Hainaut et al., 2011; Snodgrass et al., 2010; Jewitt et al., 2010; Kleyna et al., 2012).

The morphology of the dust cloud of asteroid 596 Scheila is best explained by an oblique impact (Hsieh et al., 2012; Bodewits et al., 2011; Ishiguro et al., 2011).

2201 Oljato does not show any activity and additionally it has a geometric albedo $p_V=0.43$ (Tedesco et al., 2002) while comets typically have $0.0<p_V<0.08$ (Fernandez et al., 2005) so it has also been deleted from the list.

While writing this paper a new member of the detached group was recognized, P/2012 F5 Gibbs. Although it is a AMBO it does not belong to the ABC group (Moreno et al., 2012).

The possibility that the activity of ABCs is due to a decrease of perihelion distance has not been previously demonstrated in the literature and is the objective of this paper. Additionally we will locate ABCs in an evolutionary diagram for comets.

(D) Origin in situ.

Dynamical calculations suggest that ABCs formed in situ, since models have not been able to transfer comets either from the Oort Cloud of comets or the Kuiper Belt, the two recognized reservoirs of comets (Fernandez et al., 2002; Haghighipour, 2009).

Spectra of 133P the first ABC discovered, resembles spectra of 3200 Phaeton and members of the Themis family like 62 Eratos and not spectra of traditional comets like 162P (Licandro and Campins, 2009).

(E) Rocky comets

133P is a fast rotator ($P_{ROT}$ = 3.471 h) and to ensure that the object does not fly apart, a minimum critical density is needed. It has been found that $\rho_{crit}$ = 1.33 gm/cm$^3$ for 133P, implying the existence of rocky comets (Hsieh et al. 2004) that resemble more the asteroids than traditional comets with $<\rho>=0.53\pm0.06$ gm/cm$^3$ (mean value of 19 density determinations; Ferrín, 2006).

## 3. Extinct and dormant comets

For a comet to become *extinct*, the thermal wave has to penetrate inside the nucleus up to the core and sublimate *all the ices* residing therein. It has been shown that over the age of the solar system a typical thermal wave gets exhausted at a depth of ~50 to ~150 m for known comet materials ( Prialnik and Rosenberg, 2009). Thus this is the radius of a comet for which the thermal wave has penetrated to the very center of the nucleus and presumably all volatiles have been exhausted after several perihelion passes. To be extinct, a cometary nucleus has to be small (~50 to 150 m in radius) or have a non-volatile differentiated core, which is unlikely given the density constrains.

On the other hand when the thermal wave penetrates inside a large comet nucleus, it sublimates all the volatile substances up to a certain depth that depends on thermal conductivity (composition), distance to the sun and pole orientation of the object. After several apparitions there are no more volatiles to sublimate up to that depth and the comet becomes *dormant.*

If the perihelion distance increases, the thermal wave would penetrate less, would not reach to the deep layer of ices, and the comet would become *dormant*. On the other hand, if the perihelion distance were to decrease, the thermal wave would be more intense and would penetrate deeper reaching to the ice layer, thus awakening and *rejuvenating* the comet. A new round of activity would ensue. Perihelion distances of comets change randomly due to planetary perturbations especially at resonances. An example of this is comet P/2008 R1 Garradd located very near the 8:3 resonance with Jupiter (Kleyna et al., 2012).

A decrease in perihelion distance implies a change in the received energy, ΔE, or energy enhacement, EE=ΔE, so we define a quantity $EE=1/q^2$ , shown in Figures 2 to 7 for ABCs. We find that ABCs have had perihelion decreases in recent times, thus a positive change in ΔE, and we propose that as the cause of activation. The ephemeris used and the orbital elements for this calculation are those of the Minor Planet Center (http://www.minorplanetcenter.net/iau/MPEph/MPEph.html). In the next section we consider each case individually.

## 4. Case by case study

***2P/Encke (Figure 2a).*** This is a most interesting result. It shows that even traditional comets are affected by perihelion changes. In this case it is shown that comet 2P/Encke has had and will have several episodes of ΔE = +6-7% that clearly have influenced its light curve. The next +6% ΔE will take place on 2047-2056. The secular light curve of this comet has been studied in detail by Ferrín (2008), but a detailed study of the influence of the EE on the light curve has not been acomplished.

***107P/Wilson-Harrington (Figure 2b).*** This comet was discovered as active on 1949. Ferrín et al. (2012) has shown that it was weakly active on 2005 and 2009. On 2009 the comet had a ΔE = +2.6%. We predict that the comet will be even more active on 2018-2022 when ΔE = +7.2%.

***133P/Elst-Pizarro = 7968 = 1979 OW7 (Figure 3a).*** This was the first ABC discovered. Activity was first detected on 1996 shortly after having an ΔE = +8,2% on 1983. The secular light curve of this object has been studied by Ferrín (2006). On 2040 the activity may cease due to a diminishing EE. We presume previous active apparitions were missed. The comet may have been active before 1900. The object may fade away in a few apparitions or completely by 2040 to return to activity by 2090.

***176P/LINEAR = 1999 RE70 = 118401 (Figure 3b).*** This object shows a huge energy enhancement of ΔE= 15.7% that assures a robust activity decreasing maybe by 2100. Activity may fade away in a few returns.

***233P/La Sagra = P/2009 WJ50 (Figure 4a).*** This object exhibits a secular decrease in ΔE = -2.5% per century and periodic energy enhancements of about 10%. Thus the activity of this object will necessarily cease in the near future.

***238P/Read = P/2005 U1 (Figure 4b).*** This object exhibits a significant ΔE = +11.9% on a periodic fashion, decreasing to zero by 2075.

***259P/2008 R1 Garrad (Figure 5a).*** A ΔE = +2.5% is sufficient to activate the object on 20080902 at $\Delta T = t - T_q = +38$ d after perihelion, where $T_q$ is the perihelion time. This object may have been active on several previous apparitions, 1912 and 1949. The comet is being rejuvenated at a rate ΔE = +0.3% per century.

***P/2010 R2 La Sagra (Figure 5b).*** This comet is being rejuvenated at a rate of ΔE = +0.7% per century. Additionally it exhibits periodic fluctuations of energy, with a ΔE = +3.0% on 2010 the year of discovery.

***300163 = 2006 VW139 (Figure 6a).*** This comet had a ΔE = +2.9% on 2006, the year of discovery. It exhibits periodic fluctuations of EE but no secular trend.

***3200 Phaeton = 1983 TB (Figure 6b).*** No measurable positive EE is seen. There is a secular decrease of ΔE = –2.0% per century that is making this comet more dormant. The object had very feeble activity on 2009 not due to water sublimation but to surface sputtering related to is passage through a reasonable dense, higher speed solar outflow stream (Battams and Watson, 2009). Jewitt and Li (2010) conclude that activity is due to the production of dust (through thermal fracture and decomposition-cracking of hydrated minerals) and to its ejection into interplanetary space (through radiation pressure sweeping and other effects).

***2011 CR42 (Figure 7a).*** The EE plot of this object is very interesting. It exhibits an increase in energy of $\Delta E$ = +36.8% in 1978 and 1982, the largest of the whole group. It would be interesting to look for 2011 CR42 on old sky images when it should have exhibited an abnormally large activity. This has not been done.

***596 Scheila (Figure 7b).*** This is a rather interesting case. Although the activity of this object has been shown to be due to an impact (Yang and Hsieh, 2011; Ishiguro et al., 2011; Bodewits et al.; 2011; Hsieh et al., 2012) the EE vs Year plot exhibits an enhancement $\Delta E$ = +2.1%, the smallest of this sample of objects. This values does not seem to be enough to start activity. Let us remember that P/2008 R1 Garrad had a $\Delta E$ = +2.5% which was sufficient to activate the object. So we are in the presence of a threshold or around $\Delta E$ =+2.3+0.2% to initiate activity.

***2201 Oljato (Figure 8a).*** This object exhibits no secular trend and a periodic $\Delta E$ = +4.6% which should have produced activity. However none has been detected. Previous objects in this data base present activity with less $\Delta E$. So either this is not a comet as indicated by its geometric albedo ($p_V$= 0.43), or volatiles have been exhausted long ago (see Table 1).

***P/2010 A2 LINEAR (Figure 8b).*** No enhancement of energy is apparent in this Figure, thus a perihelion diminution is discarded as a trigger of activity. This plot is consistent with activation by collision as reported by many (Kim et al., 2012; Hainaut et al., 2011; Snodgrass et al., 2010; Jewitt et al., 2010; Kleyna et al., 2012).

***P/2012 F5 Gibbs (Figure 9a).*** While writing this paper a new member of the detached group has been discovered, P/2012 F5 (Birtwhistle et al., CBET 2114). The very small energy enhancement found ($\Delta E$= +0.2%), the smallest of all the sample, suggests that this object was not activated by a diminution of the perihelion distance and this is in agreement with the interpretation by Moreno et al. (2012) and Stevenson et al. (2012) that the activation is collisional. Thus this is not an ABC.

***P/2012 T1 Panstarrs ( Figure 9b).*** This object is included in the Arxiv.org version of the manuscritp, but was identified too late to be included in the printer version. Moreno et al. (2013) have determined a dust mass loss of 6-25x$10^6$ kg. Assuming a dust to gas mass ratio of 0.5, then the total mass loss of dust plus gas is 1.8-7.5x$10^7$ kg. Using Equation (8) we find a water budget age WB-AGE= 4770-19900 cy, an old comet. Hsieh et al. (2013) also did photometry of the object and were able to set an upper limit to the water production rate Q < 5x10 mol/s. However since we do not have a light curve, we can not calculate the mass loss per revolution.

In Table 2 the properties of these objects are compiled and in particular the lag for onset of activity, $\Delta t_{LAG,}$ is estimated. For a sample of 9 comets we find $\Delta t_{LAG}$ = +48±41 d.

Ferrín (2012) and Ferrín et al. (2006) present the secular light curves of ABCs 107P/Wilson-Harrinton and 133P/Elst-Pizarro. It was found that 107P starts activity at $\Delta t$ = +26 d with a maximum around $\Delta t$ = +30±5 d, while 133P starts activity at $\Delta t$ = +42 d and reaches a maximun at $\Delta t$ = +155±10 d.

Licandro et al. (2012) present the mass loss rate for ABC 2006 VW139 = 300163. They find that the activity starts shortly after perihelion and lasts for 100 days. The dust production peaks at $\Delta t$ = +50 d.

The result by Licandro et al (2012) is in excellent agreement with the calculations presented in this paper for other ABCs, and confirms *independently* that most ABCs *tend to start activity at or after perihelion* in contrast with normal comets that start activity *well before perihelion* (Ferrín, 2010).

## 5. Mathematical Model

In the previous Section we have seen that the mean time LAG of 9 objects, true ABCs, was $\Delta t_{LAG}$ = +48±41 d. This suggests that the thermal wave has to penetrate inside the dust or crust layer to reach to the ice layer and start sublimating. In this process a time delay is introduced in the thermal wave. Since the time delay is known from observations, it may be possible to estimate the depth of the dust or crust layer. There are several caveats that have to be stated. $\Delta t_{LAG}$ is greatly influenced by the coverage or monitoring of these bodies and this is reflected in the large error bar. Additionally, the pole position is generally unknown and this influences the value of $\Delta t_{LAG}$ . Thus we are really assuming that the pole is perpendicular to the orbit. In some instances the activity may have been driven by volatiles hidden by topography or heat propagating latitudinally.

We will use a mathematical model to estimate the depth of the dust or crust layer. As it approaches perihelion, the temperature of the nucleus increases and after perihelion it decreases. This behavior can be simulated by a periodic thermal wave of period P.

Under this circumstance, the heat conduction equation for a semi-infinite homogeneous medium with a bounding surface at x=0 is

$$\nabla^2 T + Q / k = ( 1/ \alpha ) \frac{\partial T}{\partial t} \qquad (1)$$

where $\alpha = k / ( \rho c )$ = thermal diffusivity [ $cm^2 / s$ ], k is the thermal conductivity [ cal $cm^{-1}$ $s^{-1}$ $^oK^{-1}$ ], c is the specific heat [ cal $gm^{-1}$ $^oK^{-1}$ ], and $\rho$ is the density [ gm $cm^{-3}$ ], and 1 Jul = 0.238 cal. Equation (1) has a periodic solution of the form

$$T (x,t) = T_m + To\, e^{-xq} \sin ( 2 \pi \Delta t\, P - x q ) \qquad (2)$$

where $q = \sqrt{\pi / \alpha.P}$ and $T_m$ is the mean temperature of the medium. Equation (2) says that the thermal wave will be damped and will develop a time lag as a function of time as it propagates deeper into the medium. In Equation (2) x.q is the wave lag, so we ask at what depth x is the wave lag equal to the observed time lag, $\Delta t_{LAG}$:

$$x . q = x ( \pi / \alpha.P )^{1/2} = 2 \pi \Delta t_{LAG} / P \qquad (3)$$

$$x = ( 4 \pi \alpha \Delta t_{LAG}^2 / P )^{1/2} \qquad (4)$$

Thanks to the square root, the determination of the depth x is very forgiving. The propagation of errors gives

$$\Delta x / x = ( \Delta P / P + \Delta \alpha / \alpha ) / 2 + \Delta t / t \qquad (5)$$

which shows that the value of x is not sensitive to the values of P or $\alpha$ . P and $\Delta t$ are known with small errors. For $\alpha$ we will take the value derived by Muller (2007) for NEAs. Colors, sizes, albedos and orbits are very similar in NEAs and cometary nuclei. Muller (2007) finds a value for the typical thermal inertia $\alpha = 300$ J $s^{-1/2}$ $^oK^{-1}$ $m^{-2}$ = 0.0071 cal $s^{-1/2}$ $^oK^{-1}$ $cm^{-2}$. This corresponds to a thermal conductivity of 0.08 W $^oK^{-1}$ $m^{-1}$ = 8000 erg $^oK^{-1}$ $cm^{-1}$ $s^{-1}$ .

For comparison Davidsson et al. (2013) find $\alpha$ = 50-200 J $s^{-1/2}$ $^oK^{-1}$ $m^{-2}$ for comet 9P/Tempel 1.

In order to apply Equation (5) we need the period P of the thermal wave. To know it we need to calibrate the black body temperature of comets vs distance relationship, something that we will do next.

## 6. Calibration of the Black Body Temperature of Comets vs heliocentric distance

Using JHKLMNQ color photometry, researchers fit a black body curve to the thermal flux to estimate a temperature for the nucleus+dust surrounding the comet. The fit is generally very satisfactory.

Data for the black body (color) temperature of comets has been compiled from the following sources: Colangeli (1999), Hanner (1984), Hanner et al. (1984), Hanner et al. (1985), Hanner and Newburn (1989), Hanner, M.S., Newburn et al. (1987), Hanner et al. (1996), Herter et al. (1987), Jewitt et al. (1982), Kelley et al. (2006), Li et al. (2007), Lisse et al. (2004), Lynch et al. (1990) , Lorenzetti et al. (1987), Ney (1975), Oishi et al. (1978), Sitko et al. (2004), Tokunaga et al. (1992), Tokunaga et al. (1987).

This data is listed in Table 3 and plotted in Figure 10. We find that the temperature follows closely the following law

$$T = 325 \pm 5\ ^oK / \sqrt{R} \qquad (6)$$

where R [AU] is the distance to the sun.

## 7. Depth of the dust or crust layer

In Section 5 and Table 2, we have calculated that the mean maximum activity time for 9 ABCs is $\Delta t_{LAG}$ = +48±41 d, and using $\alpha = 300$ J $s^{-1/2}$ $^oK^{-1}$ $m^{-2}$ and $\alpha = 50$ J $s^{-1/2}$ $^oK^{-1}$ $m^{-2}$ equation (5) gives the results listed in Table 4. Notice the small error in the determination of the thickness. The conclusion is that the dust or crust layer has a depth of several meters, 2 to 5 m. Thanks to Equation 5, this result is robust. In fact, it looks like this is the most robust result in the literature since it is based on previously unavailable *observations*.

For comparison Deep Impact images of the nucleus of comet Tempel 1 reveal several layers 1-20 m thick that parallel the surface (Thomas et al. 2007). For asteroid Gaspra evidence suggests that it is covered with a regolith a few tens to several tens of meters thick (Carr et al., 1994). Notice the uncertainty of these determinations vs the accurate result of our mathematical model.

## 8. An Evolutionary Diagram for Comets: Remaining Revolutions vs Water-Budget Age

A) Definitions

Let us define the water budget of a comet

$$WB = \sum_{T_{ON}}^{T_{OFF}} Q_{H2O}(t)\, \Delta t \qquad (7)$$

where Q is the water production rate, and the sum is made from the onset of activity to the offset of activity. The older the comet the smaller the water budget. The Water Budget is thus the total amount of water expent by the comet per apparition.

We also need to define the water-budget age, WB-AGE (Ferrín et al., 2012). The water budget age is measured in comet years (cy) and it is a proxy for age (and activity) of a comet.

$$\text{WB-AGE [cy]} = 3.58\ \text{E+11/ WB [kg]} \qquad (8)$$

The constant is chosen so that comet 28P/Neujmin 1 has WB-AGE = 100 cy. The smaller the WB the larger the WB-AGE.

It is possible to calculate the thickness of the layer lost per apparition using the formula

$$\Delta r = (\delta + 1)\ WB / 4 \pi r^2 \rho \qquad (9)$$

where r is the radius and $\rho$ the density. We need the dust to gas mass ratio, $\delta$. Recent results by Sanzovo et al. (2009) show that $0.1 < \delta < 1.0$. Thus we will explore 2 possible scenarios with $\delta = 0.1$ and $\delta = 1.0$. The density is given by $\rho = \Delta M/\Delta V$. $\Delta V$, the volume removed, is given by $\Delta V = 4\pi r^2 \Delta r$. And $\Delta M$, the mass removed, is given by $\Delta M_{H2O} + \Delta M_{DUST} = WB\ (1 + \Delta M_{DUST}/WB)$. For the density we are going to take a value of 530 $kg/m^3$ which is the mean of 21 determinations compiled by Ferrín (2006). The number of remaining revolutions is then

$$RR = r / \Delta r = 4 \pi r^3 \rho / ( \delta + 1 )\ WB \qquad (10)$$

Notice the cubic dependence on r, which implies that smaller comet sublimate much faster. If the comet were made of pure ice, $\Delta r$ would remain constant as can be seen from the following argument. The energy captured from the Sun depends on the cross section of the nucleus, $\pi r_N^2$, on the Bond Albedo, $A_B$, and on the solar constant, S. A constant fraction of this energy is used to sublimate a volume of the comet $K\ 4\pi\ r_N^2\ \Delta r_N$. Equating both energies we find that $\Delta r_N$ should be a constant:

$$(1- A_B)\ S\ \pi r_N^2 = K\ 4\pi\ r_N^2\ \Delta r_N \qquad (11)$$

$$\Delta r_N = (1- A_B)\ S / 4K$$

and $r/\Delta r$ would tend to zero as the comet sublimates away. However if the comet contained much dust, part of it would remain on the surface, $\Delta r$ would tend to zero due to soffocation and $r/\Delta r$ would tend to infinity. Thus, sublimating away comets tend to zero. Soffocating comets tend to infinity (confirmation Figure 13).

B) Comets 2P and 103P

Knight and Schleicher (2012) have measured the water production rate of comet 103P and plotted those values for apparitions 1991, 1997 and 2010-11. This diagram modified from their Figure 2, is presented in Figure 12. (Ferrín, 2012) has calculated the WB for comet 2P. Applying the above methodology, it is possible to see the time evolution of comets 2P and 103P.

The resulting values of $RR = r / \Delta r$ are compiled in Table 5 for 14 comets at one epoch (two epochs for comets 2P/Encke and 103P). When comets in Table 5 are plotted in the Remaining Revolutions, RR, vs Water-budget Age, WB-AGE, we obtain Figure 13. This diagram encompasses 7 orders of magnitude in the vertical axis and 7 orders of magnitude in the horizontal axis. ABCs lie nearby in the upper right hand corner of the diagram, as expected. *Thus this is an evolutionary diagram for comets.* Figure 13 is reminiscent of the Herprung-Russell diagram for stars.

C) ABC 2006 VW139

The JPL site on "physical data" for solar system objects found at http://ssd.jpl.nasa.gov gives the H magnitude of ABC 2006 VW139 as H=16.1. If we assume a geometric albedo of $p_v$=0.05 then the radius is 1.8 km and the diameter 3.6 km. If we assume a density of 0.53 g/cm$^3$, mean value of 21 density estimates of comets (Ferrín, 2006), it is possible to calculate the thickness of the upper layer lost by the comet during that apparition using equation (8). Licandro et al. (2012) have calculated a dust mass loss of 2x10(6) kg for ABC 2006 VW139.

From Table 5 we see that comet 2006 VW139 lost 1.9E-4 cm in the 2009 apparition. Since the radius of this comet is 1800 m, the ratio $r_N$ / $\Delta r_N$ = 9.7E6 for $\delta$ = 1 (Figure 13). This is the most extreme object in the upper right hand corner of that diagram. This calculation implies that the comet has been choked by dust, producing suffocation. This value also suggests that the comet has reached the end of its evolutionary path, since it is no longer capable of removing significant amounts of the surface layer. Table 4 shows that this might be due to a layer of dust or crust 1.9 m in depth.

D) 3200 Phaethon

Li and Jewitt (2013) have found recurrent activity in 3200 Phaeton, starting half a day after perihelion and lasting for 3-5 days. This result is in agreement with the conclusions presented at the end of Section 4, and confirms *independently* that most ABCs *tend to start activity at or after perihelion* in contrast with normal comets that start activity *well before perihelion* (Ferrín, 2010).

The observations are surprising in view of the result presented in Figure 6 that this object has a negative EE, strengthening the conclusion by Li and Jewitt that the activity can not be driven by sublimation. They find that the comet lost a dust mass M ~4x10^8 $a_{mm}$ kg, where $a_{mm}$ is the grain radius expressed in mm. The object can be modelled as a sphere of 2.5 km radius. Assuming $a_{mm}$ = 1, 0.1, and 0.01 mm and a density of 0.53 g/cm$^3$ as before, we find the $\Delta$r and r/$\Delta$r values using Equations (9) and (10). We can find an age if we expand our definition of water-budget age to a mass-loss age that includes any mass loss not related to water sublimation. Then

$$\Delta r = 4x10\text{^}8\ a_{mm} / 4\ \pi\ r^2\ \rho \tag{12}$$

$$ML\text{-}AGE\ [cy] = 3.58\ E{+}11/\ ML\ [kg] \tag{13}$$

The results listed in Table 5 and plotted in Figure 15, show that 3200 Phaeton belongs to the graveyard of comets.

**9. The Graveyard**

The RR vs WB-AGE diagram covers the region of the graveyard of comets, those objects with 1000 cy < WB-AGE. Five members belong to the graveyard, 107P, 133P, 2006 VW139, D/1891 W1 Blanpain and 3200 Phaeton.

Thus we propose that the asteroidal belt contains an enormous graveyard of ancient dormant and extinct rocky comets, that turn on (are rejuvenated), in response to a diminution of their perihelion distance.

**10. Conclusions**

We have arrived at the following conclusions:

(1) The asteroidal belt contains an enormous graveyard of *ancient dormant* and *extinct rocky* comets that turn on mostly after perihelion. The bodies are *ancient* because these objects were active long time ago and the ice layer has since then receded deep inside their bodies. The objects are *dormant* because their orbits have remained stable for millions of years, after which the ice layer has receded inside the body beyond reach of the thermal wave, making the object dormant. The comet is *extinct* if the radius of the body is less than the penetration of the thermal wave. Then, after many apparitions, all ices up to the center of the body have been exhausted and no ice remains to be activated even if the perihelion distance were to decrease substantially. They are *rocky* because it has been shown that 133P has a density of 1.33 $gm/cm^3$, greater than water ice.

(2) We define the *"detached group"* as those objects that exhibit cometary characteristics (sublimate water ice) and have aphelion distances Q < 4.5 AU. The detached group contains all the ABCs traditionally recognized, plus a few other members not traditionally recognized like 2P and 107P. With the above definition there are 12 members of the ABC group: 2P,

107P, 133P, 176P, 233P, 238P, C/2008 R1, C/2010 R2, 2011 CR42, 3200, 300163 = 2006 VW139 and P/2012 T1 Panstarrs. While writing this paper a new member of the detached group was recognized: P/2012 F5 Gibbs but it does not belong to the ABC group because it seems it was activated by a collision.

(3) In the literature a common reason for activity is interplanetary collisions. We present an alternative explanation. Once active, objects are sublimating ices except for P/2010 A2, 596 Scheila and P/2012 F5 Gibbs that have exhibited dust tails due to collisions and 3200 Phaethon activated by solar wind sputtering.

(4) *Comets rejuvenate*, and we traced the origin of activity to a diminution of their perihelion distances, q, caused by planetary perturbations, a hypothesis that has not been previously explored in the literature.

(5) As a consequence of this model, *we predict* that ABCs will start activity (turn on) *mostly after* perihelion. Two comets have been shown to become active after perihelion: 107P ($\Delta t$ = + 38 d) (Ferrín et al., 2012), 133P ($\Delta t$ = + 42 d ) (Ferrín, 2006). Most of the other objects listed in Table 2, were discovered after perihelion (mean $\Delta t_{LAG}$ = +48±41 d). Recently Licandro et al. (2012) have studied ABC 2006 VW139 and found that it starts activity after perihelion and the activity peaks at $\Delta t_{LAG}$ = +50 d. Li and Jewitt (2013) have found that 3200 Phaeton also starts activity after perihelion. These two papers validate our conclusions *independently*.

These results support the idea that the thermal wave has to penetrate deep into the surface to activate a deeply buried layer of ice. Thus there should be a stratified layer of ices inside the nucleus under a crust or layer of dust. Sublimating away comets do not have a layer of dust or crust.

(6) We find secular trends:

- The following ABCs are being rejuvenated: 133P, 176P, P/2008 R1, P/2009 R2 (4/11 = 36%).
- The following ABCs are becoming more dormant: 233P, 3200 (2/11 = 18%).
- The following ABCs do not show any secular trend but exhibit periodic variations of EE: 2P, 107P, 238P, 2006 CR41, 2206 VW139 (5/11 = 45%).
- The following objects are active due to collisions and thus are not true ABCs : P/2010 A2, 596 Scheila, P/2012 F5 Gibbs (3/14=21%).
- The following object is not an ABC due to its large albedo: 2201 Oljato.

(7) We have calibrated the black body (color) temperature of comets vs perihelion distance, R. We find T = 325±5 ºK/√R.

(8) Using a mathematical model we calculate the thickness of the crust or dust layer on comet nuclei. We find a thickness of 2.0±0.5 m for comet 107P, 4.7±1.2 m for comet 133P and 1.9±0.5 m for a sample of 9 comets. Notice the small error due to the mathematical model.

(9) We have located four ABCs in an evolutionary diagram of Remaining Revolutions, RR, vs Water-budget Age, WB-AGE. This diagram encompasses 7 orders of magnitude in the vertical axis and 7 orders of magnitude in the horizontal axis. ABCs lie together in the upper right hand corner of the diagram, as expected from physical arguments.

(10) The RR vs WB-AGE diagram defines the region of the graveyard of comets, those objects with 1000 cy < WB-AGE. Five members belong to the graveyard, 107P, 133P, 2006 VW139, D/1891 W1 Blanpain and 3200 Phaeton. Thus we propose that the asteroidal belt contains an enormous graveyard of ancient dormant and extinct rocky comets, that turn on (are rejuvenated), in response to a diminution of their perihelion distance.

(11) One important question remain unanswered: When did the cometary activity in the main belt peaked and when did it end?

(12) Since the ABCs are objects rejuvenated from a graveyard of dormant and dead comets, we suggest to call them “The Lazarus Comets”.

**Acknowledgements**

We thank CODI of the University of Antioquia for their support through project E01592. We thank an anonymous referee for many useful suggestions that improved the scientific quality of this paper, and who preferred to remain anonymous.

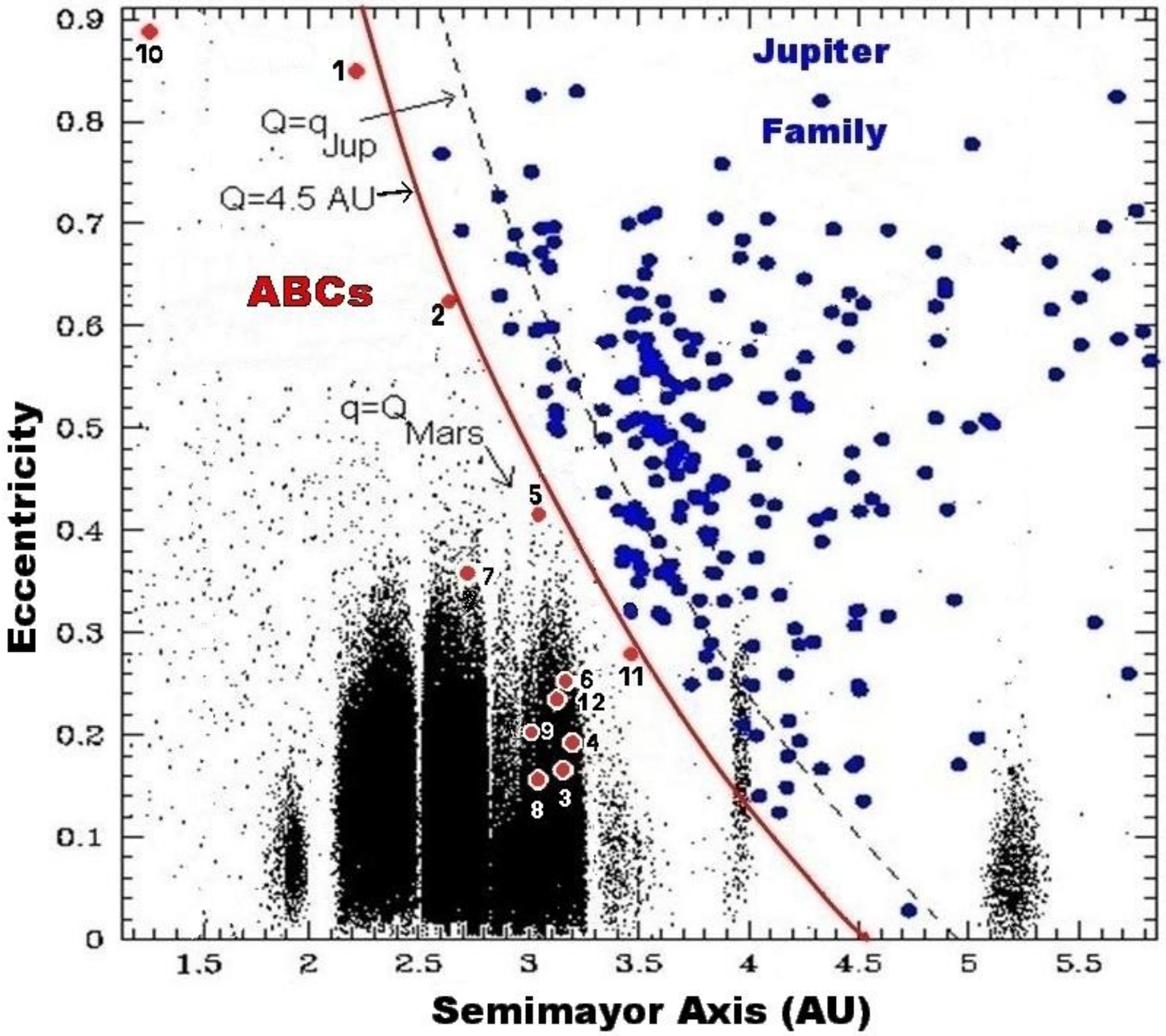

***Figure 1.*** *The location of ABCs and Jupiter family comets in the eccentricity vs semi-major axis diagram. The ABC numbers correspond to the listing in Table 1. Object 10 at the extreme upper left corner ( 3200 Phaeton ) was included by Jewitt and Li (2010) in the ABC list, opening the door to include objects 1, 2, 5, and 11. The line Q=4.5 AU, where Q is the aphelion distance, separates ABCs from the Jupiter Family. 1 = 2P, 2 = 107P, 3 = 133P, 4 = 176P, 5 = 233P, 6 = 238P, 7 = P/2008 R1, 8 = P/2010 R2, 9 = 300163, 10 = 3200, 11 = 2011 CR42, 12 = P/2012 T1. It is found that all the objects lie at Q < 4.5 AU. The dashed line has Q=q(JUP). The full line has Q=4.5 AU. Notice how sharply the continuous line separates two population of comets. Does 4.5 AU have a physical interpretation?*

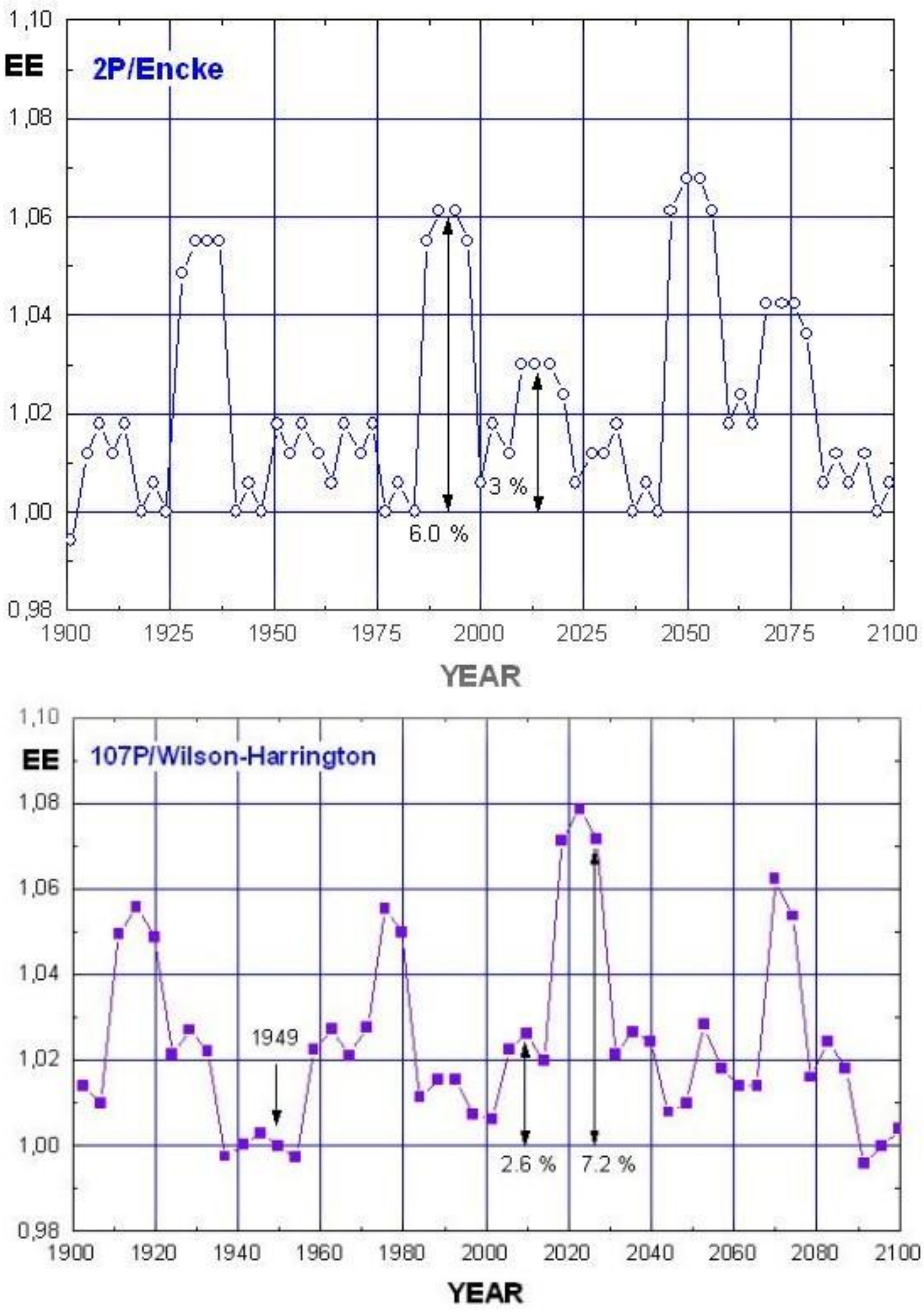

***Figure 2***. *200 year diagram of Energy Enhancement, EE, vs Year. See text for a description of individual objects. The energy enhancement is shown. These plots should be useful to predict the activity of these objects in the future.*

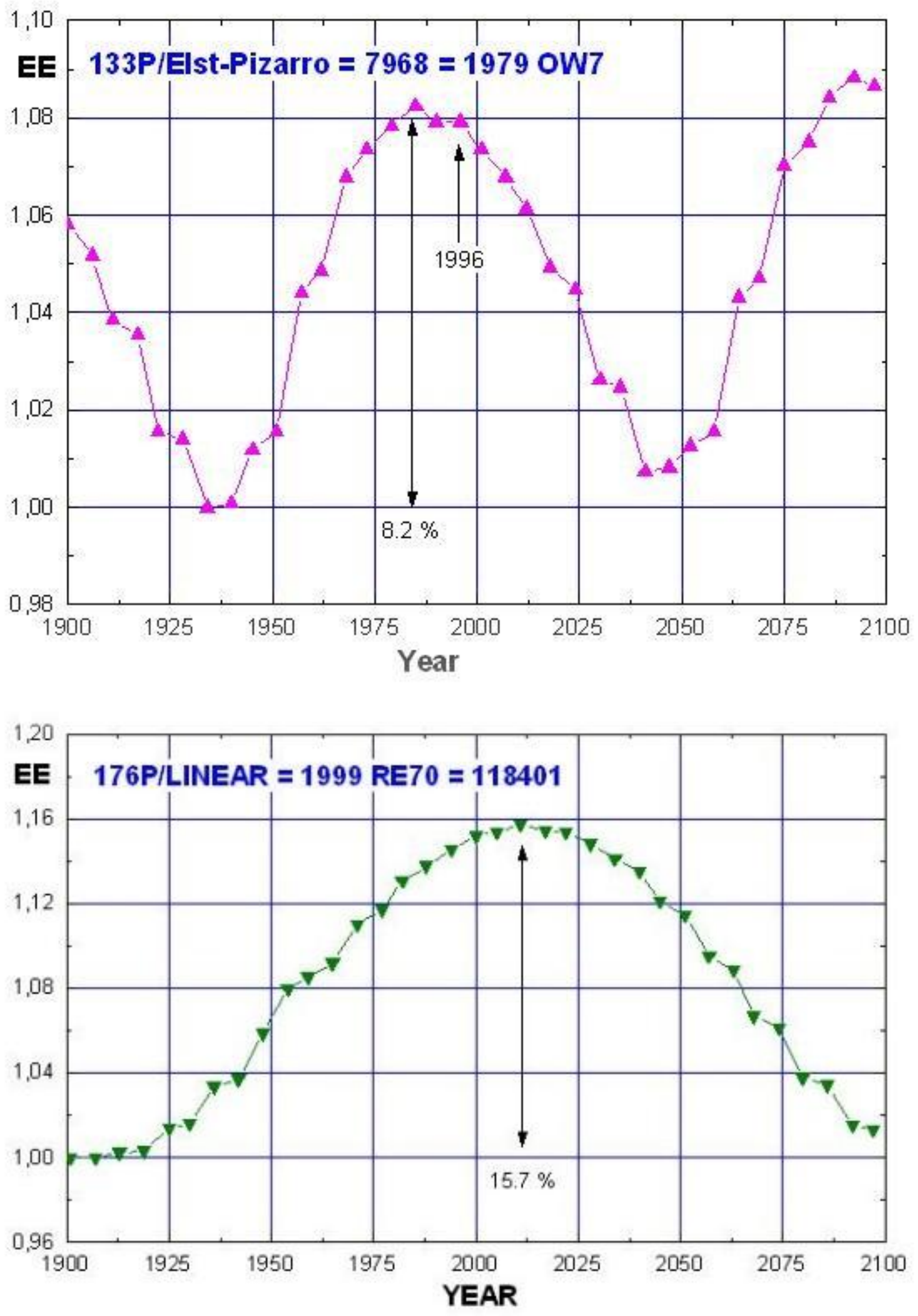

***Figure 3.*** *200 year diagram of Energy Enhancement, EE, vs Year. See text for a description of individual objects. The year of discovery of the object is indicated as well as the energy enhancement.*

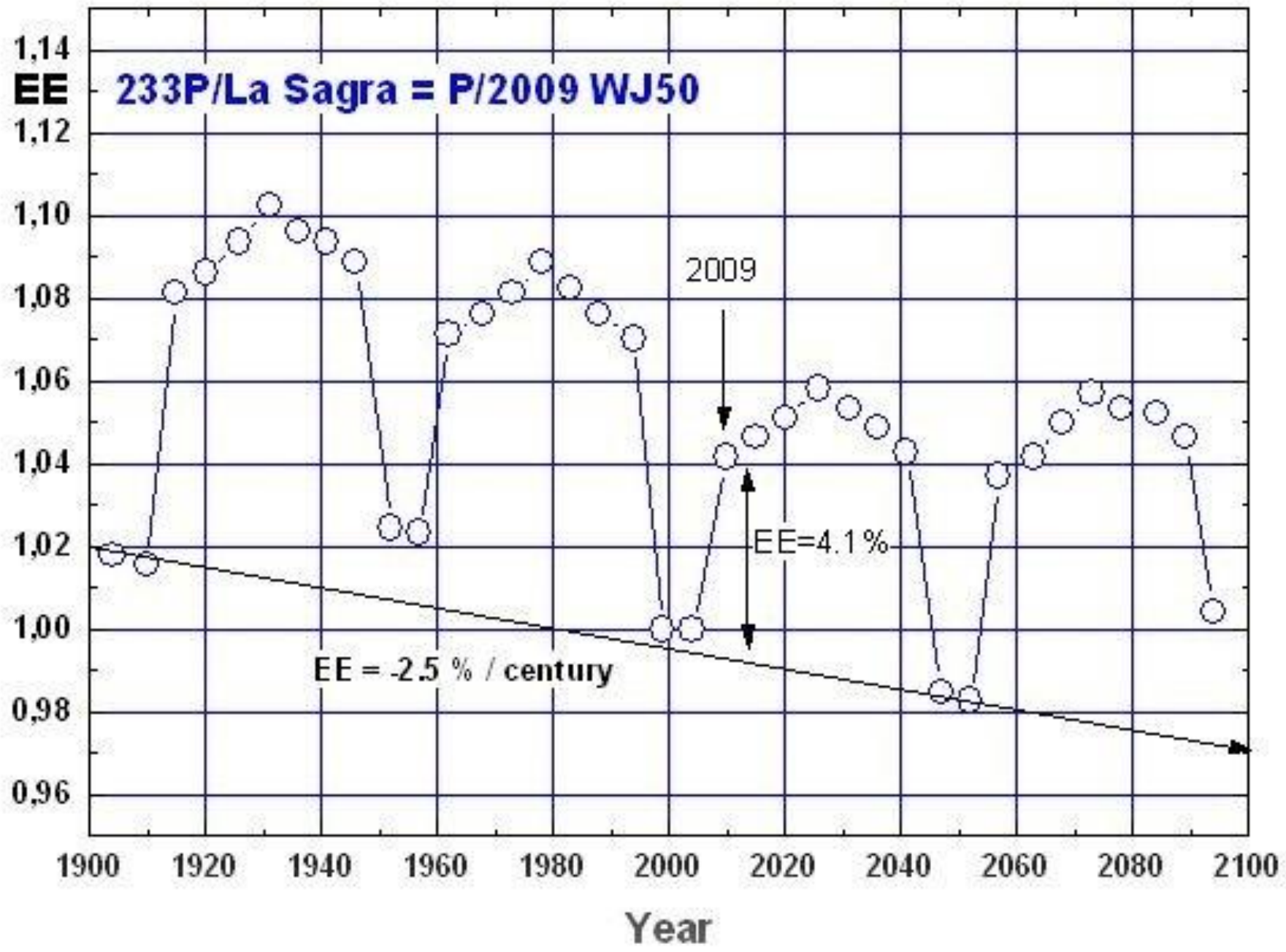

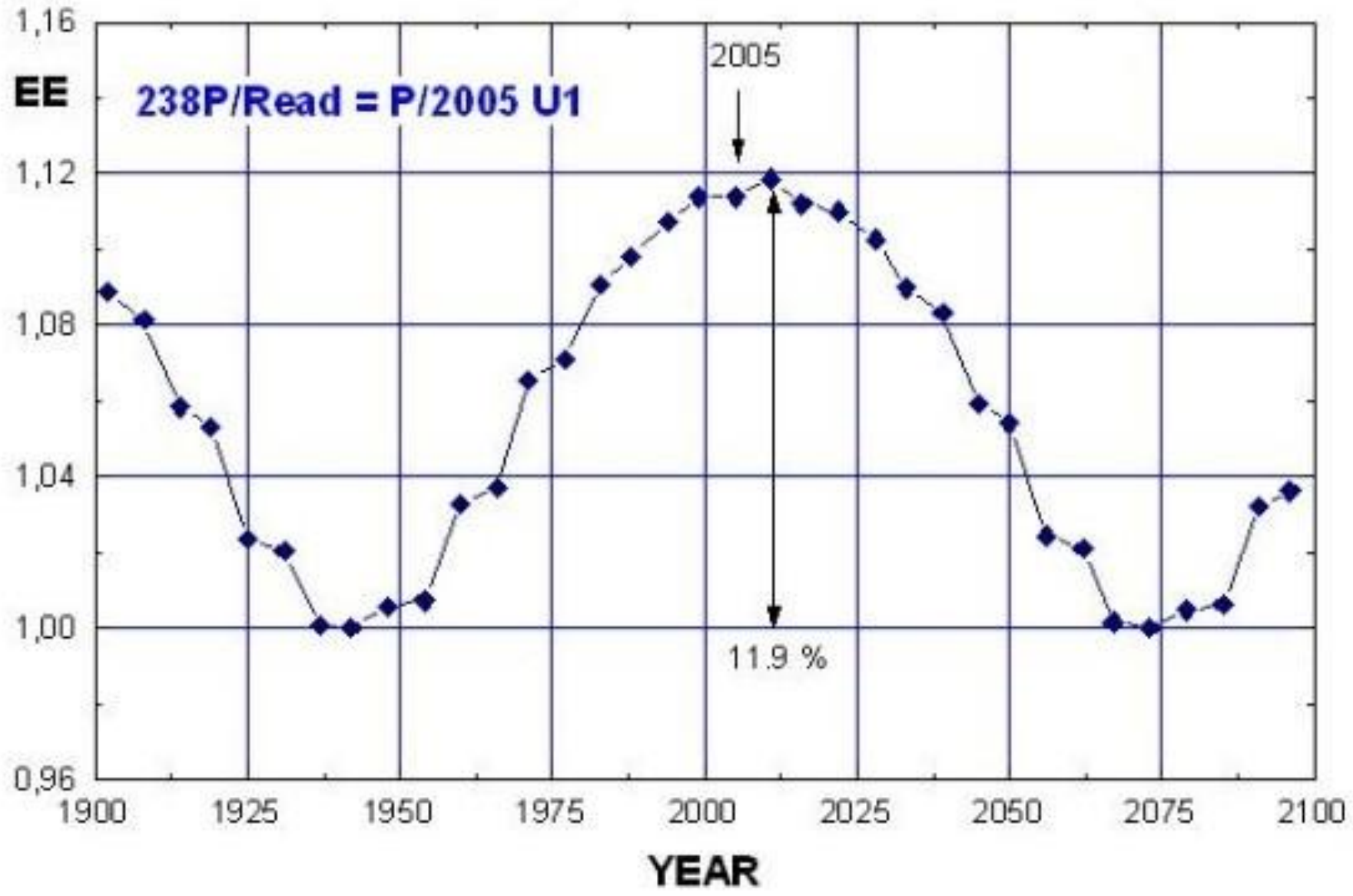

***Figure 4.*** *200 year diagram of Energy Enhancement, EE, vs Year. See text for a description of individual objects. The year of discovery of the object is indicated as well as the energy enhancement.*

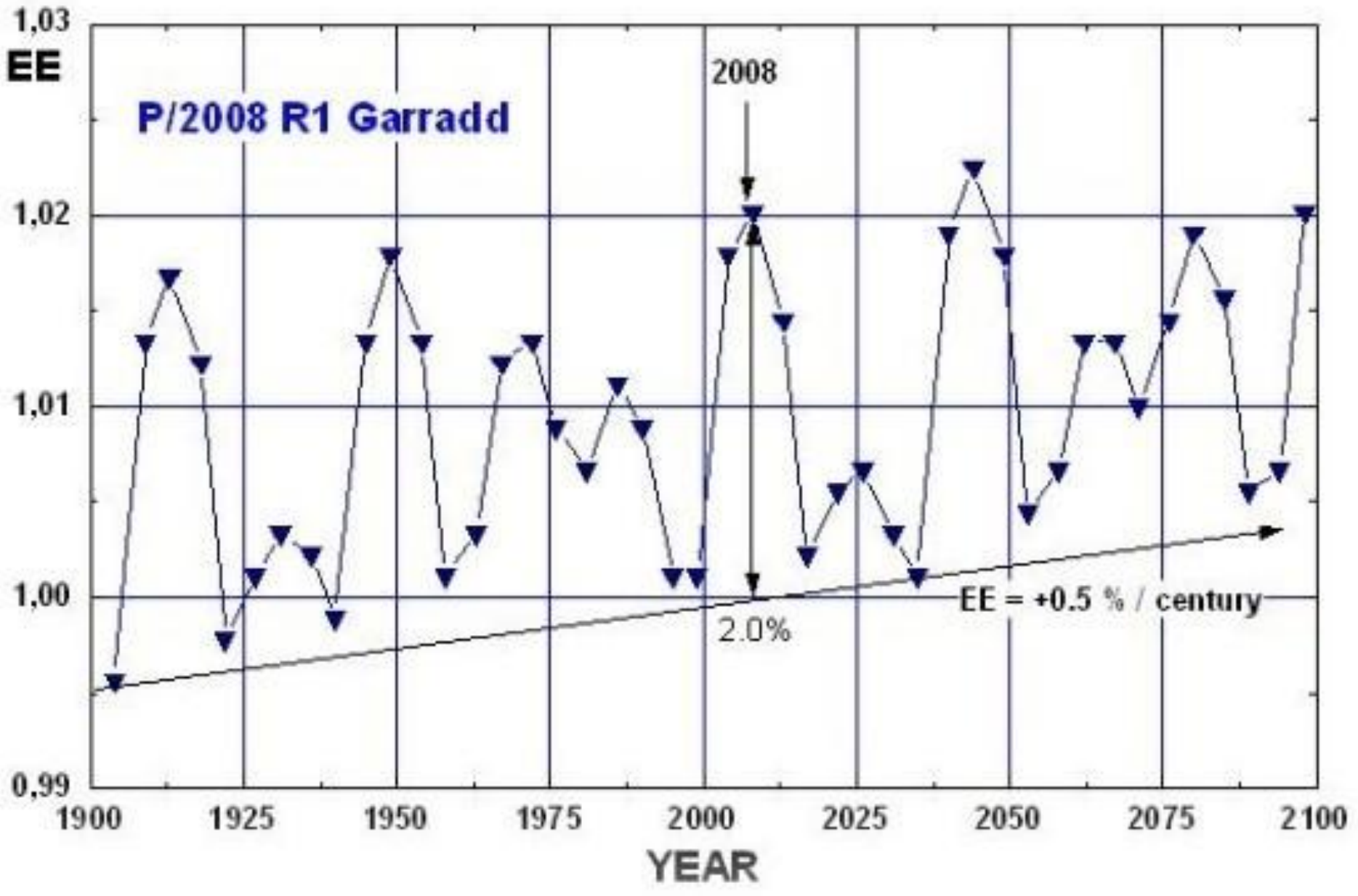

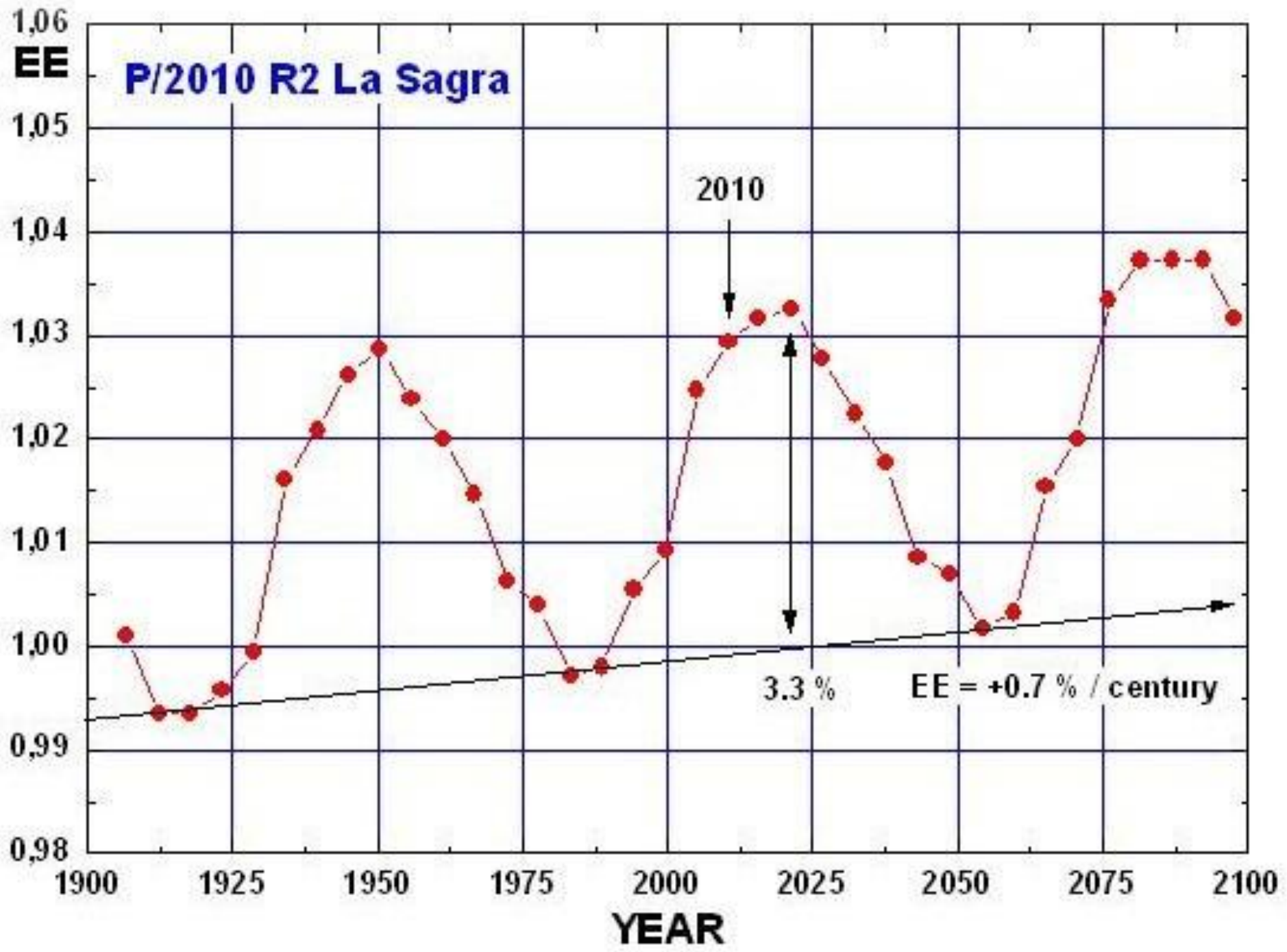

***Figure 5.*** *200 year diagram of Energy Enhancement, EE, vs Year. See text for a description of individual objects. The year of discovery of the object is indicated as well as the energy enhancement.*

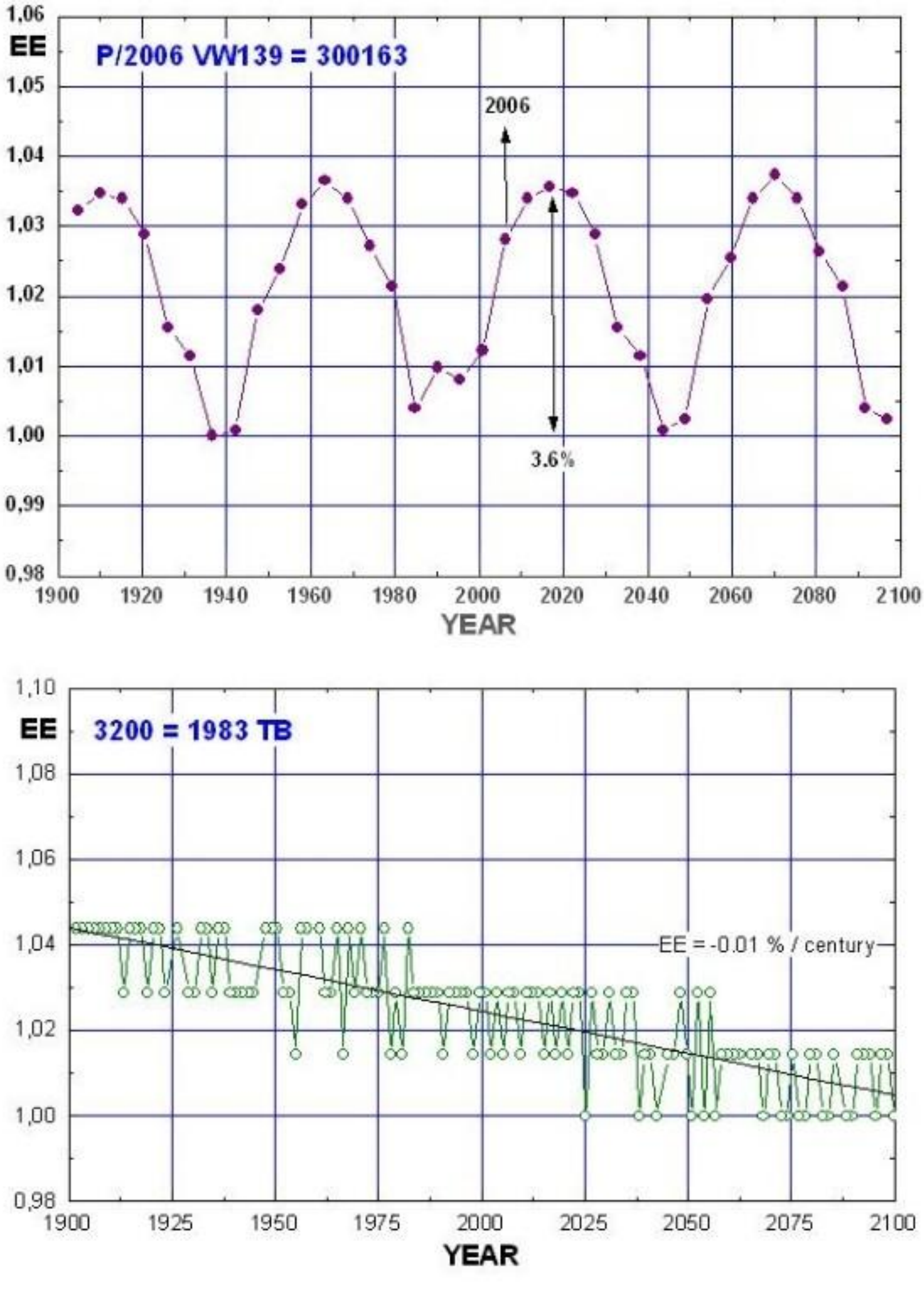

***Figure 6.*** *200 year diagram of Energy Enhancement, EE, vs Year. See text for a description of individual objects. The year of discovery of the object is indicated as well as the energy enhancement. In spite of the fact that 3200 Phaeton is becoming more dormant Li and Jewitt (2013) have found that the comet is active for a short period of time after perihelion. This plot supports their conclusion that activity is not driven by sublimation.*

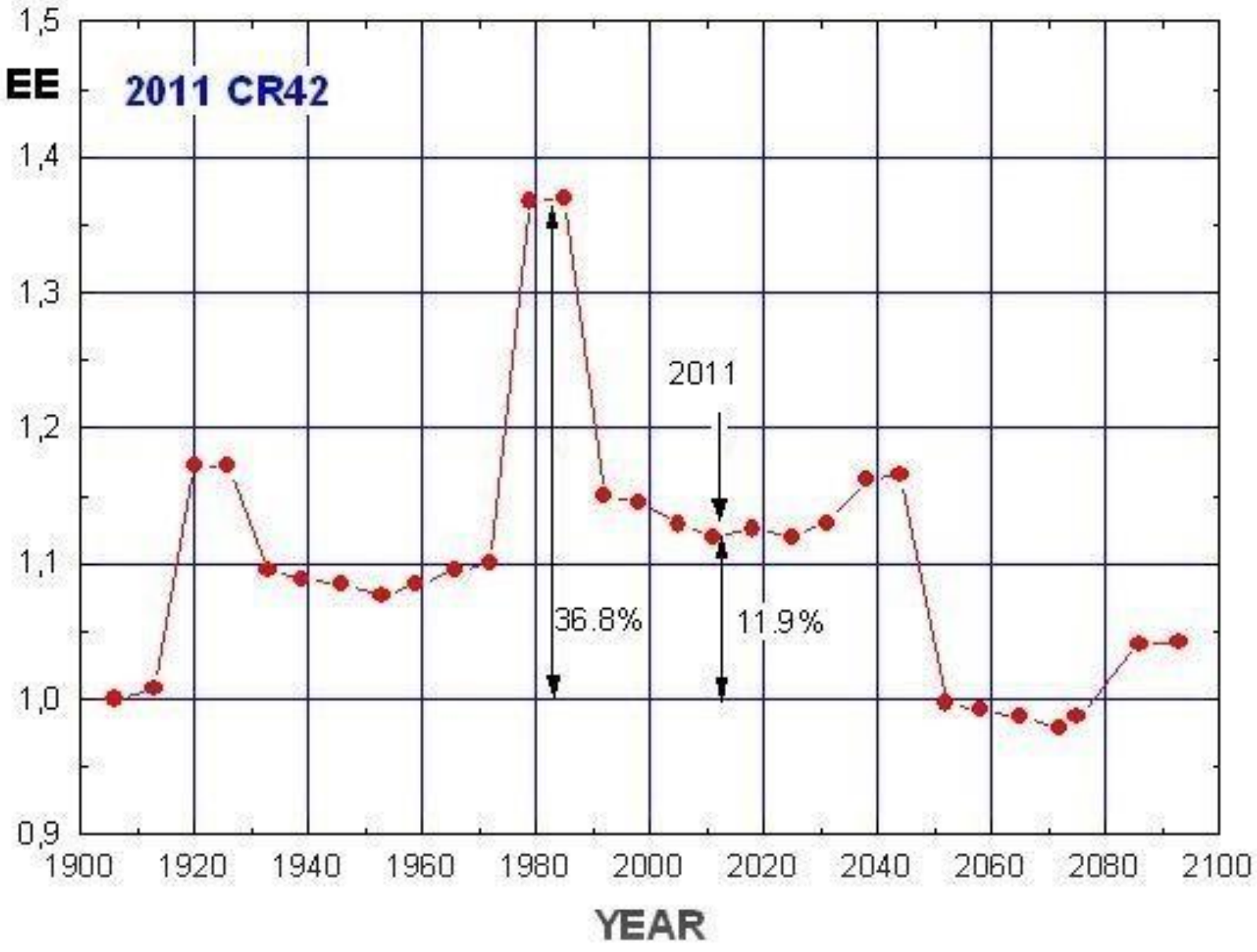

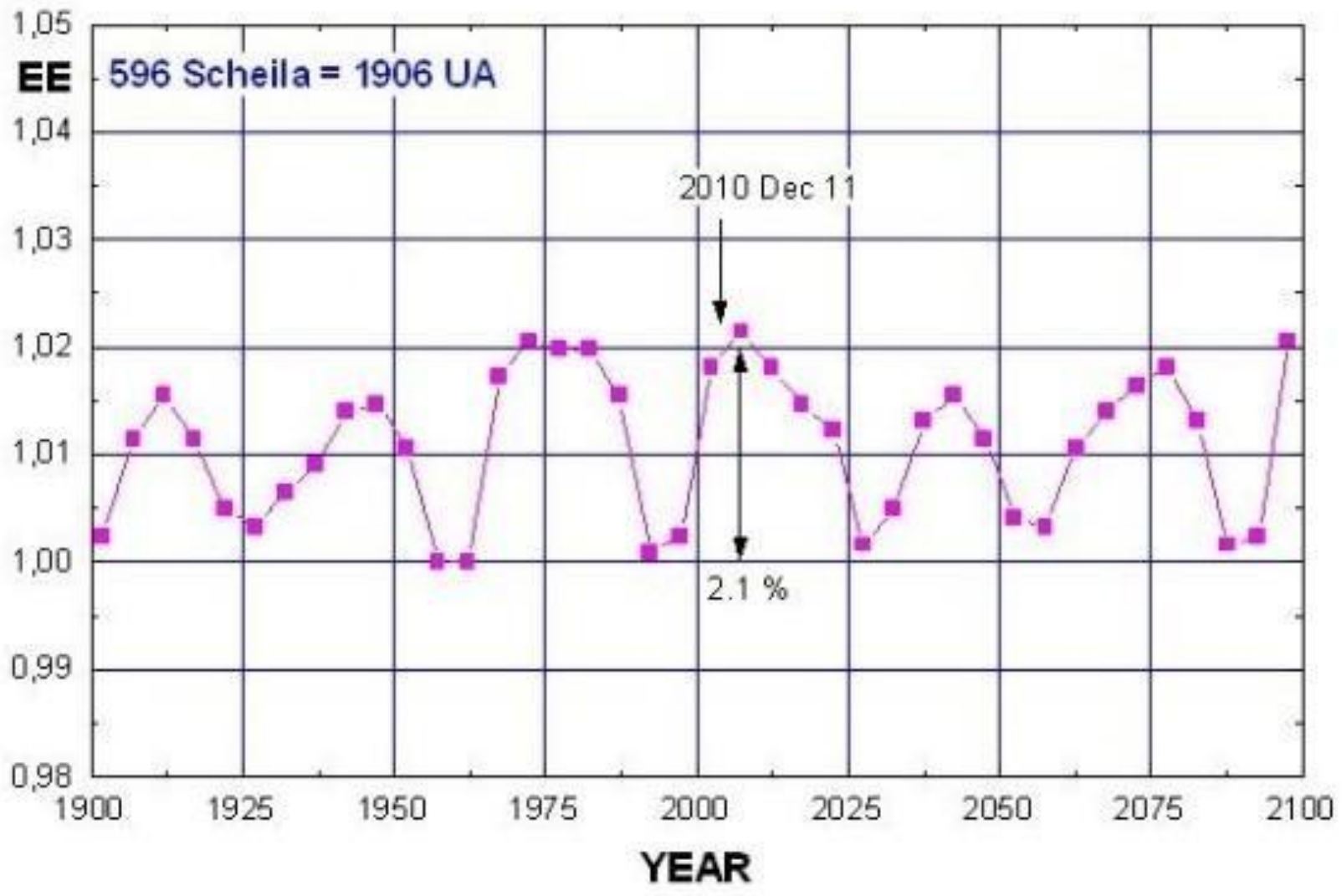

***Figure 7.*** *200 year diagram of Energy Enhancement, EE, vs Year. See text for a description of individual objects. The year of discovery of the object is indicated as well as the energy enhancement. Neither 2011 CR42 nor 596 Scheila show a secular trend.*

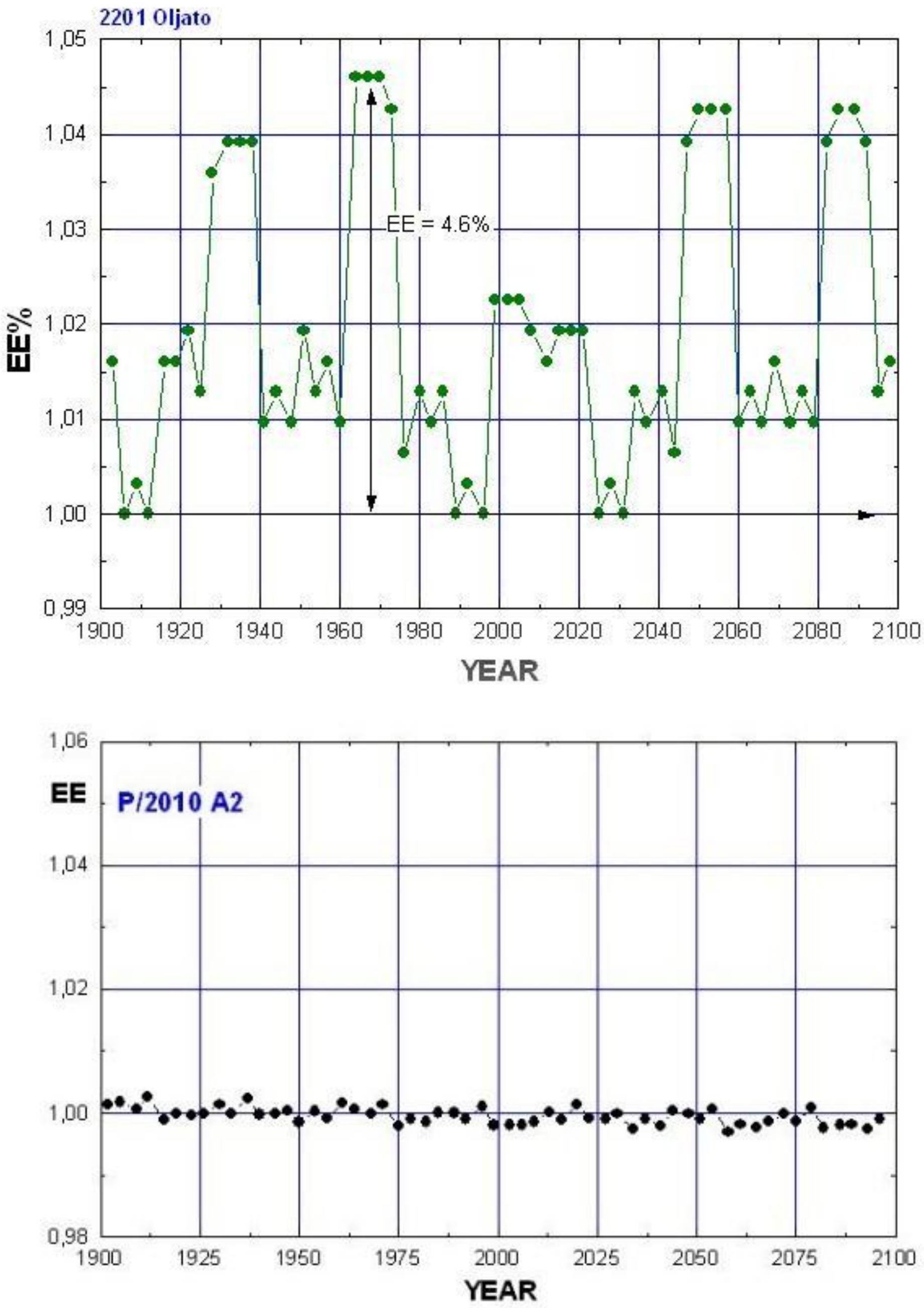

***Figure 8.*** *200 year diagram of Energy Enhancement, EE, vs Year. See text for a description of individual objects. The year of discovery of the object is indicated as well as the energy enhancement. 2201 has a huge energy enhancement. However it does not get active. Neither 2201 nor P/2010 A2 show any secular trend. This result in the case of P/2010 A2 is consistent with a collisional activation.*

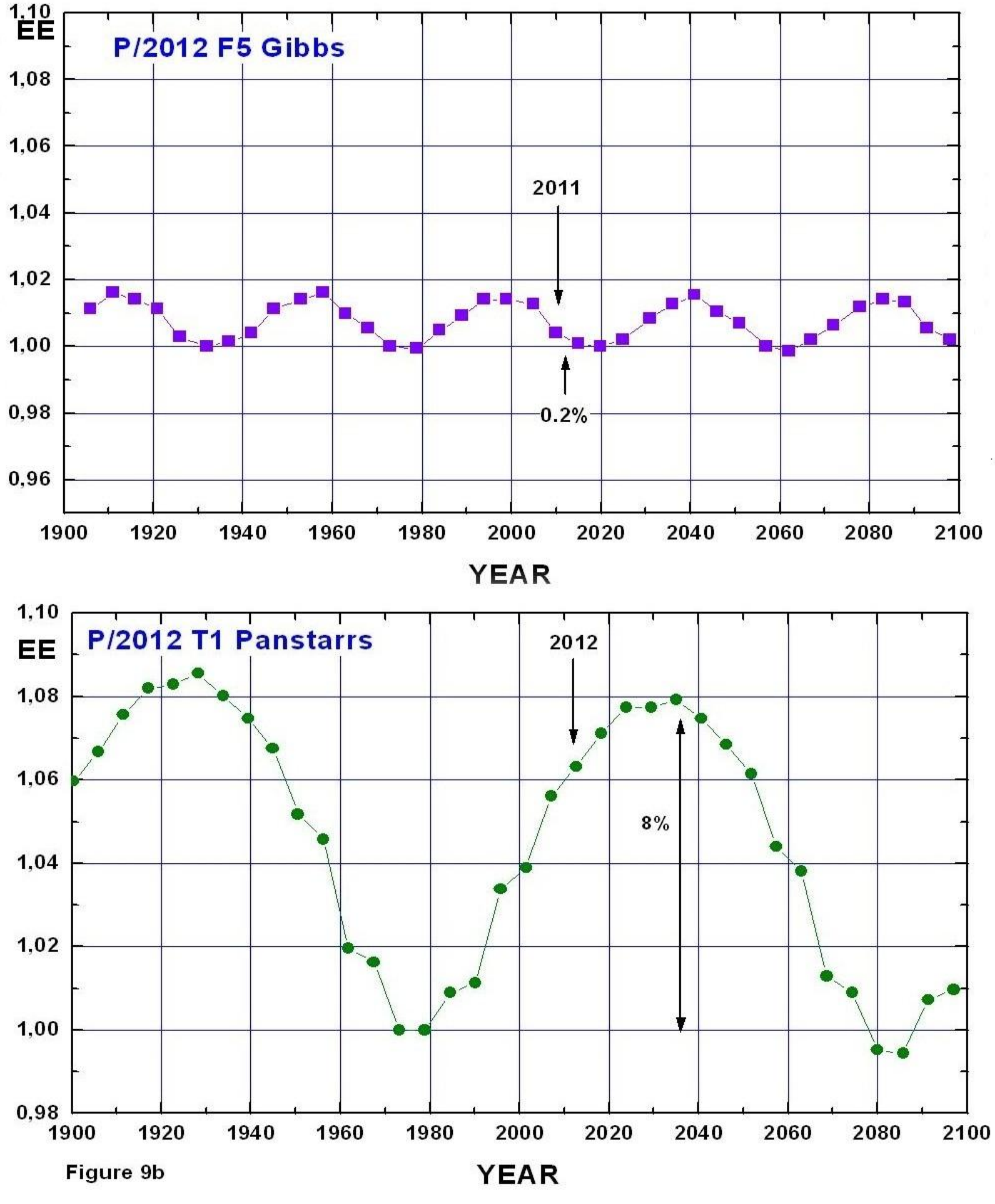

***Figure 9.*** *P/2012 F5 exhibits a very small EE of 0.2% that we consider insufficient to initiate activity and it is well below the threshold ΔE =+2.3+0.2% stablished in Section 4. This fact combined with an activation time of Δt = +462 d, supports the hypothesis of Moreno et al.(2012) that activation of this object was due to a collisional event. On the other hand, P/2012 T1 Panstarrs exhibits an energy enhancement of 8%, sufficient to show strong activity.*

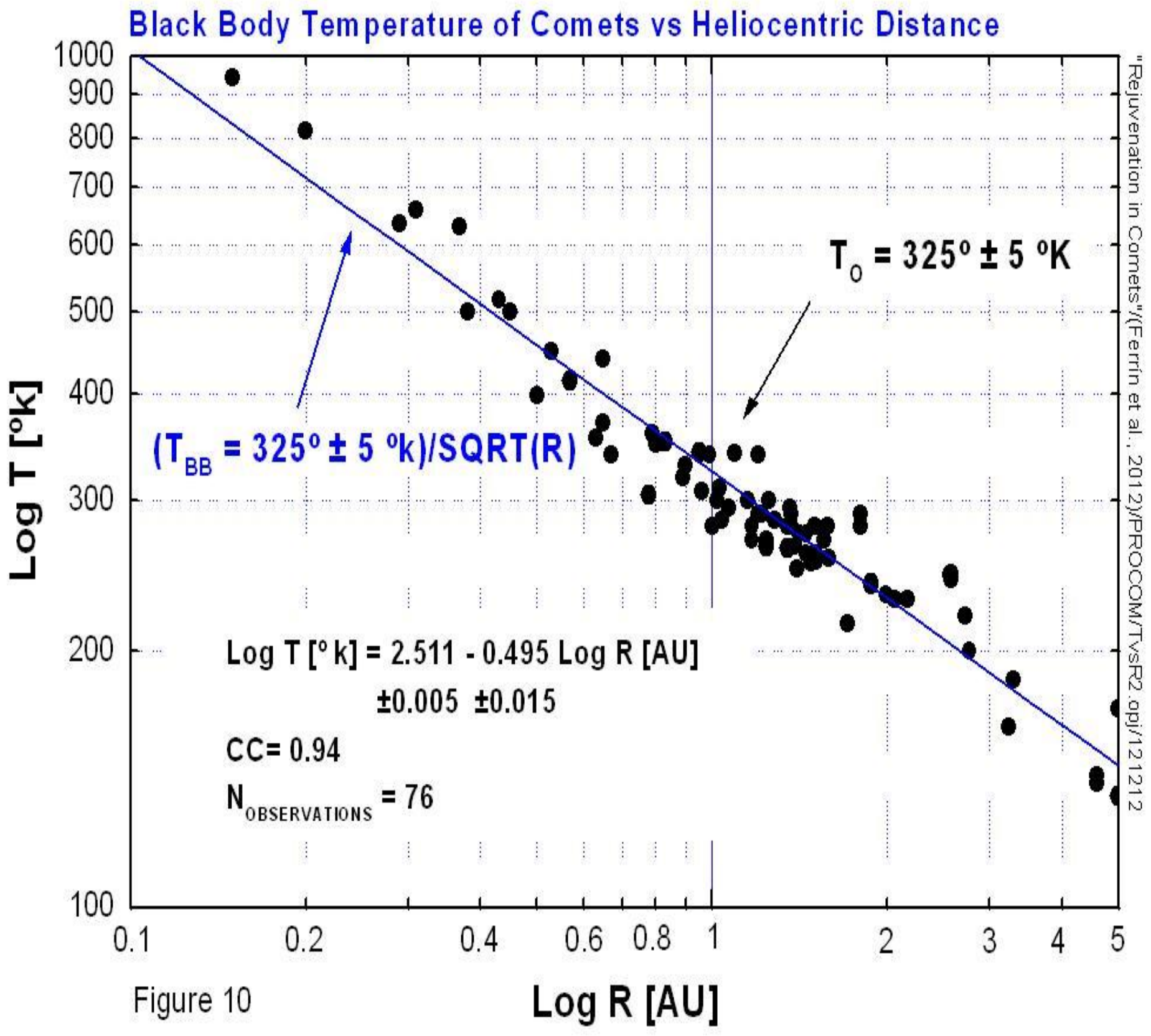

***Figure 10.*** *Calibration of the Black Body (color) Temperature vs heliocentric distance, R, for comets regardless of class. The fit by least squares gives the value of the function,* $T = 325 \pm 5 ºK / \sqrt{R}$ .

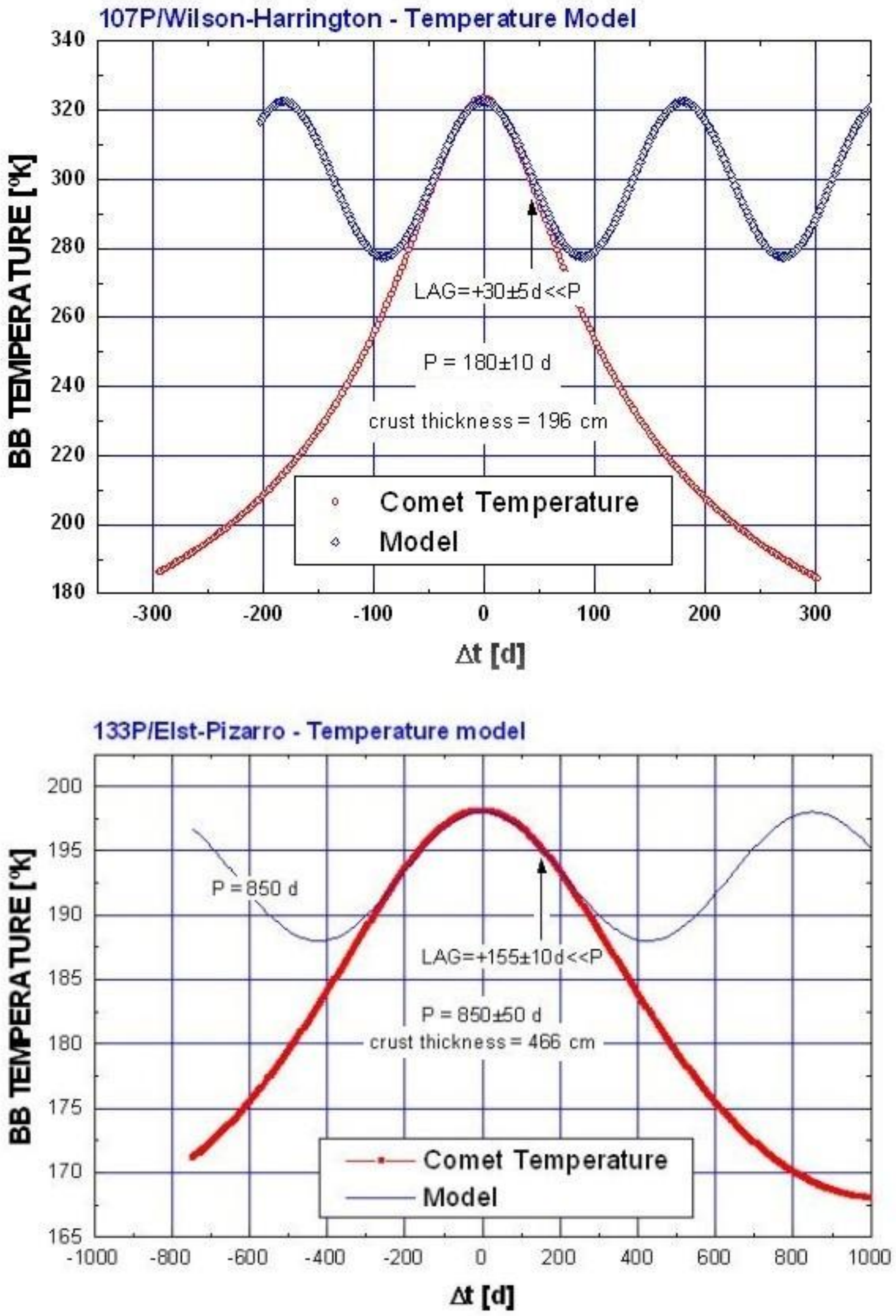

***Figure 11.*** *The bell shaped line describes the thermal wave on a comet based on an observational calibration of the black body color temperatures for comets presented in Figure 10. The periodic line shows the fit with a sinusoid of period P. Since LAG << P the fit is excellent. The vertical arrow pointing up is the time of onset of maximum sublimation. In both cases this takes place after perihelion, in contrast with normal comets that get activated well before perihelion (Ferrín, 2010).*

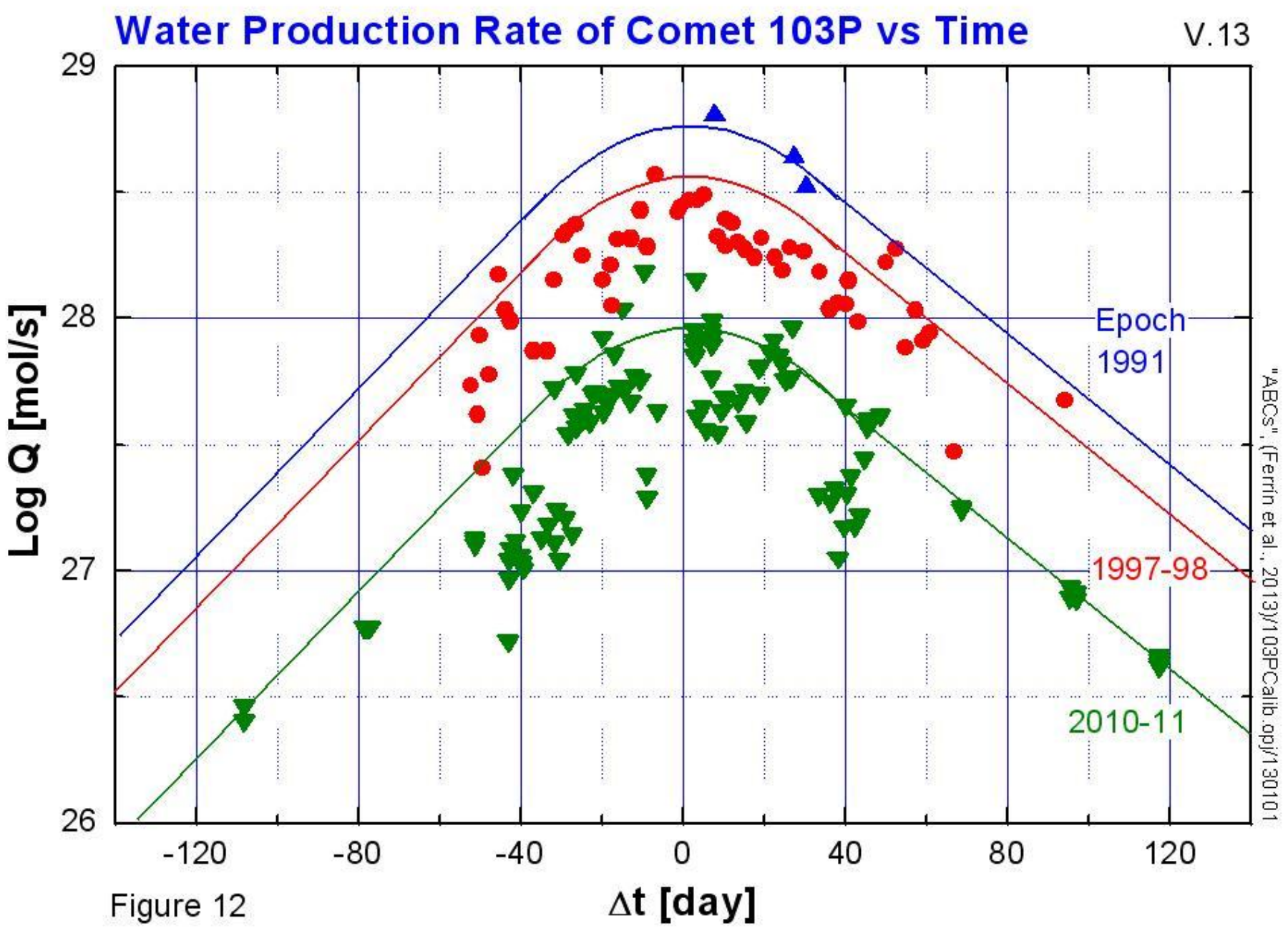

*__Figure 12.__ The water production rate of comet 103P vs time and epoch. This diagram has been modified from Figure 2 of Knight and Schleicher (2012). It shows a decreasing production rate as a function of epoch. The shape of the curve is about the same for all epochs. The envelope of the distribution has been chosen as a definition of the production rate because some observations are affected by insufficient integration time, insufficient subtraction of background or insufficiente measuring apperture. These effects affect the observed values downward. Additionally, the upper part of the distribution (the envelope) is rather sharp and well defined, while the anti-envelope is diffuse and show larger scatter. There is no physical effect that can bring observations upward. So we conclude that the envelope of the observations is the correct interpretation of the production rate plot. The decrease in production rate from 1991 to 2010-11 is by a factor of 6.3, a huge value. Thus this comet shows a rapid evolution rate and this is also reflected in the remaining revolutions vs water budget age, shown in Figure 13.*

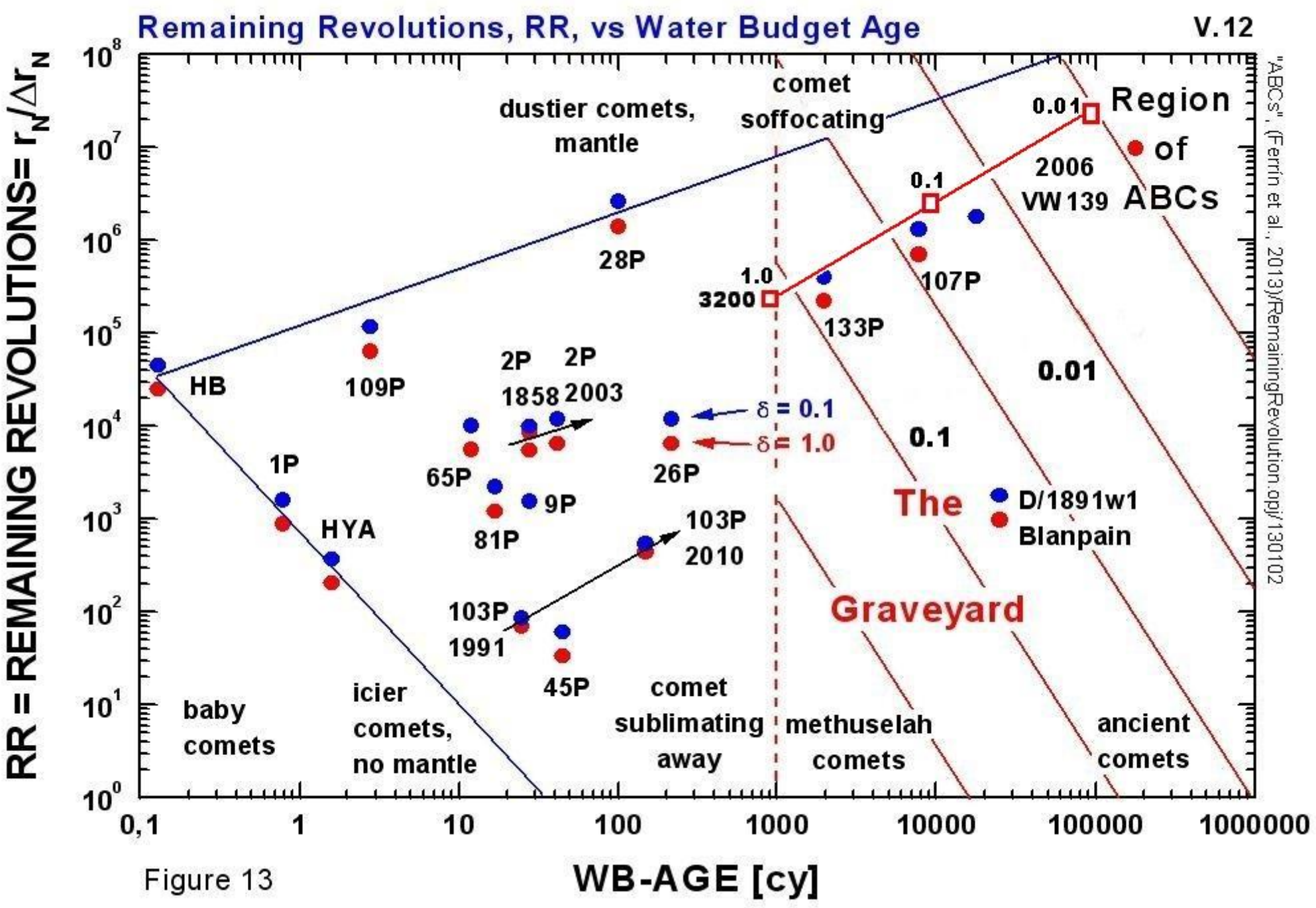

***Figure 13.*** *Remaining Revolutions vs Water Budget Age. This diagram is notorious in that the vertical and the horizontal axis each span a range of 7 orders of magnitudes. This is the range of activity of this sample of comets in the two parameter phase space. If the comet were made of pure ice, Δr (the layer removed by apparition) would remain constant and r/Δr would tend to zero as the comet sublimates away. If the comet contained much dust, part of it would remain on the surface, Δr would tend to zero and r/Δr would tend to infinity. Comet 2P/Encke shows evolution of the parameters between the 1858 and the 2003 apparitions. Comet 103P shows evolution between the 1991 and the 2010-11 apparitions. Thus this is an evolutionary diagram. Comets evolve from the left side toward the right side. If they move up they are choked by a dust crust. If they move down they sublimate away. 2006 VW139 is the most extreme object in the upper right hand corner. The location of a comet is not sensitive to the dust to gas mass ratio. Two comets are sublimating away, 103P and 45P. Five comets belong to the graveyard in this definition, 107P, 133P, D/1891W1 Blanpain, 2006 VW139 and 3200 Phaeton. The location of 3200 is plotted for three diameters of the dust particles, 1.0 mm, 0.1 m and 0.01 mm. It is not surprising that ABCs occupy the upper right hand corner of the diagram. This is expected on physical grounds. The diagram says that they are old (large WB-AGE) and that they have a substantial dust or crust layer (large r/Δr ). The depth of this layer is calculated in the text with the aid of a mathematical model.*

Table 1. Physical properties of the detached group[1] , DG.
AMBO = DG = ABC + CA

| # CLASS | Object | q AU | Q AU | Dia km | $P_{ROT}$ h | $p_V$ | a AU | e | i ° | Tiss | $V_N$ |
|---|---|---|---|---|---|---|---|---|---|---|---|
| ABC | 2P= =P/1786 B1 | 0.33 | 4.09 | 5.1 | 22.2 | 0.04 | 2.21 | 0.85 | 11.8 | 3.02 | ---- |
| ABC | 107P=4015 =P/1949 W1 | 0.99 | 4.28 | 3.5 | 6.3 | 0.06 | 2.64 | 0.62 | 2.78 | 3.08 | ---- |
| ABC | 133P=7968 P/1996 N2 | 2.64 | 3.68 | 3.8[4] | 3.47 | 0.05[4] | 3.16 | 0.17 | 1.4 | 3.18 | 15.9 |
| ABC | 176P= 118401 1999RE70 | 2.58 | 3.81 | 4.0[4] | 22.2 | 0.06[4] | 3.20 | 0.19 | 0.2 | 3.17 | 15.1 |
| ABC | 233P/La Sagra P/2009= WJ50 | 1.79 | 2.27 | --- | ---- | ---- | 3.03 | 0.41 | 11.3 | ---- | ---- |
| ABC | 238P/Read P/2005 U1 | 2.36 | 3.97 | 0.8 | ---- | ---- | 3.17 | 0.25 | 1.3 | 3.15 | 17.7 |
| ABC | 259P/2008 R1 Garradd | 1.79 | 3.66 | 0.44 0.60[8] | ---- | ---- | 2.73 | 0.34 | 15.9 | 3.22 | 20.8[7] |
| ABC | P/2010 R2 La Sagra | 2.62 | 3.58 | 1.6[5] | ---- | 0.03[5] | 3.09 | 0.15 | 21.4 | 3.10 | 15.5 |
| ABC | 300163 = 2006 VW139 | 2.44 | 3.66 | 3.6 | ---- | ---- | 3.05 | 0.20 | 3.2 | 3.20 | ---- |
| ABC | 3200=1983 TB Phaeton | 0.14 | 2.40 | 7 5.13 | 3.6 | 0.10 | 1.27 | 0.89 | 22.2 | 4.5 | ---- |
| ABC | 2011 CR42 | 2.53 | 4.49 | ---- | ---- | ---- | 3.51 | 0.28 | 8.45 | ---- | ---- |
| ABC | P/2012 T1 | 2.42 | 3.68 | ---- | ---- | ---- | 3.05 | 0.21 | 11.4 | ---- | ---- |
| CA | P/2012 F5 | 2.87 | 3.12 | 4.2[6] | ---- | ---- | 3.00 | 0.04 | 9.7 | 3.19 | ---- |
| CA | 596 Scheila[2] | 3.40 | 3.43 | 113 | 15.8 | 0.04 | 2.93 | 0.17 | 14.7 | 3.21 | 8.9 |
| CA | P/2010A2 | 1.99 | 2.59 | 0.12 | ---- | ---- | 2.29 | 0.13 | 5.3 | 3.58 | 21.3 |
| -- | 2201 Oljato[3] | 0.63 | 3.71 | 1.8 | 26 | 0.43 | 2.17 | 0.71 | 2.52 | 2.52 | 17.8 |

1. Detached means detached from Jupiter, Q<4.5AU
2. 596, P/20100 A2 and P/2012 F5 were active due to an impact and thus are not true ABCs
3. 2201 Oljato has an albedo pv=0.43 and thus does not belong to the class of comets.
4. Hsieh et al. (2009)
5. Bauer et al. (2011)
6. Stevenson et al. (2012)
7. MacLenan, E.M., Hsieh, H. (2012)
8. Kleyna et al. (2012).

Table 2. Discovery of activity and turn on of activity
of active comets. $\Delta t_{LAG}$ = time from perihelion

| Object | CLASS | Discovery of activity = T2= turn on time | REF | Tperihelion = T1 | $\Delta t_{LAG}$ = T2-T1[c] |
|---|---|---|---|---|---|
| 3200[a] Phaeton | ABC | 20090617 | IAUC 9054 | 20090620 | +0.5 d |
| 133P P/1996 N2 | ABC | 19960807 | IAUC 6456 | 19960421 | +108 d |
| 176P= 118401 1999 RE70 | ABC | 20051126 | IAUC 8704 | 20051019 | +38 d |
| 133P/La Sagra P/2009 WJ50 2005 JR71 | ABC | 20100206 | IAUC 9119 IAUC 9117 | 20100312 | -34 d |
| 238P= P/2005 U1 Read | ABC | 20051024 | IAUC 8624 | 20050728 | +88 d |
| P/2008 R1 Garradd | ABC | 20080902 | IAUC 8969 | 20080725 | +39 d |
| P/2010 R2 La Sagra | ABC | 20100915 | CBET 2459 | 20100626 | +81 d |
| 300163 = 2006 VW139 | ABC | 2011 08 30 | CBET2920 | 20110718 | +43 d |
| 2011 CR42[b] | ABC | 20110305 | CBET 2823 | 20111130 | -272 d |
| 107P/W-H[b] | ABC | 19491119 | Bowell. CIAU 5585 | 20091022 | +26 d |
| 133P/E-P[b] | ABC | 19960530 | CIAU 8847 | 19960418 | +42 d |
| P/2012 T1 | ABC | 20121006 | CBET3252 | 20121122 | -45 d |
| P/2010 A2 | CA | 20100106 | CBET 2114 | 20100507 | +121 d |
| P/2012 F5 | CA | 20110701 | Moreno et al. (2012) | 20100327 | +462 d |
| 596 Scheila[a] | CA | 20101211 | CBET 2583 | 20120518 | -523 d |

a - 596 was activated by a collision near aphelion. 3200 exhibited feeble activity at perihelion caused by solar wind sputtering. 3200 is currently a dormant ABC.

b - 107P and 133P are true comets being active at several apparitions, and we suspect 2011 CR42 it is too due to its activation well before perihelion.

c- The mean value and standard error of the mean of this sample of comets excluding the active comets and the collisionally activated objects 596, P/2012 F5, 107, 133, 2011 CR42, P/2010 A2, is $\Delta t$ = T2-T1= +48±41 d, which implies that most ABCs are activated after perihelion. If all pole orientations were perpendicular to the orbits, then the observed time lag can be interpreted in terms of a thermal wave penetrating into the nucleus and activating deeper layers of ice. This is done in Section 5.

Table 3. Black Body (color) Temperature of Comets

| Comet | R (AU) | T(°k) | References |
|---|---|---|---|
| 1P/Halley | 0.79 | 360 | Sitko et al. (2004) |
| | 0.80 | 350 | Hanner et al. (1987) |
| | 0.89 | 320 | Hanner et al. (1987) |
| | 0.90 | 330 | Hanner et al. (1987) |
| | 0.96 | 308 | Hanner et al. (1987) |
| | 1.07 | 294 | Hanner et al. (1987) |
| | 1.01 | 222 | Reach et a. (2013) |
| | 1.25 | 300 | Sitko et al. (2004) |
| | 1.28 | 285 | Herter et al. (1987) |
| | 1.88 | 240 | Lorenzetti et al.(1987) |
| | 1.99 | 233 | Tokunaga et al. (1987) |
| 2P/Encke | 1.17 | 270 | Lisse et al. (2004) |
| | 0.38 | 500 | Kelley et al. (2006) |
| | 1.09 | 341 | Kelley et al. (2006) |
| | 2.56 | 246 | Kelley et al. (2006) |
| | 2.57 | 243 | Kelley et al. (2006) |
| | 2.57 | 247 | Kelley et al. (2006) |
| 4P/Faye | 1.51 | 260 | Sitko et al. (2004) |
| | 1.59 | 257 | Hanner et al. (1996) |
| 10P/Tempel 2 | 1.39 | 266 | Tokunaga et al. (1992) |
| 19P/Borrelly | 1.80 | 280 | Li et al. (2006) |
| | 1.45 | 275 | Hanner et al. (1996) |
| | 1.45 | 260 | Sitko et al. (2004) |
| 23P/Brorsen-Metcalf | 0.50 | 399 | Lynch et al. (1990) |
| 24P/Schaumasse | 1.24 | 265 | Hanner et al. (1996) |
| | 1.24 | 270 | Sitko et al. (2004) |
| 26P/Grigg-Skje | 1.02 | 300 | Hanner et al. (1984) |
| | 1.15 | 300 | Hanner et al. (1984) |
| | 1.36 | 294 | Hanner et al. (1984) |
| 38P/Stephan-Oterma | 1.58 | 280 | Hanner et al. (1984) |
| 55P Tempel-Tuttle | 1.03 | 310 | Sitko et al. (2004) |
| 65P/Gunn | 2.77 | 200 | Hanner et al. (1984) |
| 67P/Churyumov-Gerasimenko | 1.37 | 290 | Hanner et al. (1985) |
| | 1.35 | 280 | Hanner et al. (1985) |
| | 1.50 | 280 | Hanner et al. (1985) |
| | 1.48 | 254 | Hanner et al. (1985) |
| | 1.88 | 239 | Hanner et al. (1985) |
| | 4.98 | 180 | Kelley et al. (2006) |
| | 4.98 | 135 | Kelley et al. (2006) |
| 69P/Taylor | 2.17 | 230 | Sitko et al. (2004) |
| 103P/Hartley 2 | 1.04 | 285 | Colangeli et al. (1999) |
| 126P/IRAS | 1.71 | 215 | Lisse et al. (2004) |
| C/1973 E1 Kohoutek | 0.15 | 944 | Ney (1975) |
| | 0.2 | 818 | Ney (1975) |
| | 0.29 | 636 | Ney (1975) |
| | 0.31 | 660 | Sitko et al. (2004) |
| | 0.43 | 517 | Ney (1975) |
| | 0.45 | 500 | Ney (1975) |
| | 0.57 | 415 | Ney (1975) |
| | 0.65 | 440 | Ney (1975) |
| | 0.83 | 352 | Ney (1975) |
| | 0.95 | 342 | Ney (1975) |
| C/1975 V1-A West | 0.368 | 630 | Oishi et al. (1978) |
| | 0.530 | 450 | Oishi et al. (1978) |
| C/1980 E1 Bowell | 4.60 | 140 | Jewitt et al. (1982) |
| C/1980 Y2 Panther | 1.80 | 290 | Jewitt et al. (1982) |
| C/1983 O1 Cernis | 3.3 | 185 | Hanner (1984) |
| C/1983 H1 IRAS-Araki-Alckock | 1.00 | 280 | Sitko et al. (2004) |
| C/1986 P1 Wilson | 1.20 | 340 | Sitko et al. (2004) |
| | 1.35 | 264 | Hanner & Newburn (1989) |
| C/1987 P1 Bradfield | 0.99 | 340 | Sitko et al. (2004) |
| C/1989 Q1 Okazaki-Levy-Rudenko | 0.65 | 370 | Sitko et al. (2004) |
| C/1989 X1 Austin | 0.78 | 305 | Sitko et al. (2004) |
| C/1990 K1 Levy | 1.56 | 270 | Sitko et al. (2004) |
| | 1.51 | 255 | Sitko et al. (2004) |
| C/1993 A1 Mueller | 2.06 | 230 | Sitko et al. (2004) |
| C/1995 O1 Hale-Bopp | 2.73 | 220 | Sitko et al. (2004) |
| C/1999 T1 McNaught-Hartley | 1.40 | 250 | Sitko et al. (2004) |
| | 1.41 | 275 | Sitko et al. (2004) |
| C/2001 HT50 LINEAR-NEAT | 3.24 | 163 | Kelley et al. (2006) |
| | 4.60 | 143 | Kelley et al. (2006) |
| C/2002 O4 Honing | 1.37 | 280 | Sitko et al. (2004) |
| C/2002 T1 Juels-Holvorcem | 1.22 | 290 | Sitko et al. (2004) |
| C/2002X5Kudo-Fujikawa | 0.67 | 340 | Sitko et al. (2004) |
| | 0.63 | 355 | Sitko et al. (2004) |
| C/2002 V1 NEAT | 1.20 | 290 | Sitko et al. (2004) |
| | 1.17 | 280 | Sitko et al. (2004) |

N(observations) = 76
N(comets) = 34

Table 4. Results: Thickness of the dust or crust layer.

| Sample | $\Delta t_{LAG}$ [d] | P [d] | Thickness[cm] $\alpha = 300$ | Thickness[cm] $\alpha = 50$ |
|---|---|---|---|---|
| 9 Comets | 48±41 | 515 | 185±45 | 75±20 |
| 107P | 30±5 | 180±10 | 196±51 | 80±26 |
| 133P | 155±10 | 850±50 | 466±123 | 190±62 |
| 2006 VW139 | 50±5 | 515 | 193±47 | 79±24 |

Table 5. Water budget, WB, water budget age, WBAGE, Remaining Revolutions, RR=$r_N/\Delta r_N$=RR(δ) were δ = M(dust)/M(gas). WB-AGE [cy] = 3.58 E+11/ WB [kg].

| Comet | WB [kg] | WB-AGE [cy] | WB --------% WB(1P) | $r_N$ [km] | $\Delta r_N$ [m] δ = 1 | RR δ=0.1 | RR δ=1 |
|---|---|---|---|---|---|---|---|
| HB | 2.67 E+12 | 0.13 | 590 | 27 | 1.1 | 44600 | 24550 |
| 1P | 4.51 E+11 | 0.79 | 100 | 4.9 | 5.6 | 1580 | 868 |
| Hya | 2.25 E+11 | 1.6 | 50.0 | 2.4 | 12 | 370 | 204 |
| 109P | 1.29 E+11 | 2.8 | 28.6 | 13.5 | 0.21 | 115500 | 63500 |
| 65P | 3.06 E+10 | 12 | 6.8 | 3.7 | 0.67 | 10020 | 5500 |
| 19P/Borrelly | 2.17E+10 | 16 | 4.8 | 2.25 | 1.3 | 3178 | 1748 |
| 81P | 2.09 E+10 | 17 | 4.6 | 1.97 | 1.6 | 2200 | 1200 |
| 103P 1991[1] | 1.41 E+10 | 25 | 3.1 | 0.57 | 8.1 | 86<br>δ=0.02 | 70<br>δ=0.25 |
| 103P 2010-11[1] | 2.24E+09 | 159 | 0.5 | 0.57 | 1.3 | 540<br>δ=0.02 | 440<br>δ=0.25 |
| 2P2003 | 8.58 E+09 | 42 | 1.9 | 2.55 | 0.40 | 11700 | 6436 |
| 2P1858 | 1.28 E+10 | 27 | 2.8 | 3.20 | 0.75 | 15500 | 8525 |
| 9P | 1.27 E+10 | 28 | 2.8 | 2.75 | 0.50 | 9900 | 5450 |
| 45P | 7.95 E+09 | 45 | 1.8 | 0.43 | 12.9 | 60 | 33 |
| 28P | 3.58 E+09 | 100 | 0.8 | 11.5 | 0.008 | 2.6E06 | 1.4E06 |
| 26P | 1.64 E+09 | 218 | 0.4 | 1.47 | 0.02 | 11700 | 6450 |
| 133P | 1.81 E+08 | 1978 | 0.32 | 2.3 | 0.010 | 4.0E05 | 2.2E05 |
| 107P | 2.03 E+07 | 7800 | 0.004 | 1.65 | 0.002 | 1.3E6 | 7E5 |
| D/1819 W1 Bla | 1.40E+07 | 2.5E4 | 0.003 | 0.16 | 0.16 | 1770 | 970 |
| 2006 VW139 | 2.00E+06 | 1.8E5 | 4.4E-6 | 1.8 | 1.9E-4 | ------ | 9.7E6 |
| 2006 VW139 | 2.00E+07 | 1.8E4 | 4.4E-5 | 1.8 | 1.0E-3<br>δ=0.1 | 1.8E6 | ------- |
| 3200[2] | 4.00E+08 | 9E2 | 8.9E-4 | 2.5 | 9.6E-3 | ----- | 2.6E5 |
| 3200[2] | 4.00E+07 | 9E3 | 8.9E-5 | 2.5 | 9.6E-4 | ----- | 2.6E6 |
| 3200[2] | 4.00E+06 | 9E4 | 8.9E-6 | 2.5 | 9.6E-5 | ----- | 2.6E7 |

1. For comet 103P Sanzovo et al. (2010) found 0.02 < δ < 0.25
2. For comet 3200 Phaeton, Li and Jewitt (2013) find only a dust ejection. For this comet we do not calculate a water-budget age but a mass-loss age. See Section 9 for a discusion of this data.

## - THIS ADDITIONAL MATERIAL WILL NOT APPEAR IN THE PRINTED VERSION. IT IS ONLY FOR ILLUSTRATION

Caption. The main belt of asteroids is represented at two epochs, now and at ancient times.

**Above.** In the old paradigm there was no cometary activity going on in the main belt.

**Middle.** In the new paradigm there is some cometary activity currently going on in the main belt.

**Bottom.** In this investigation the situation in the main belt is clarified. There was a vigorous cometary activity in ancient times, but now it has subsided to residual activity. In consequence the main belt *must contain a graveyard of comets*.

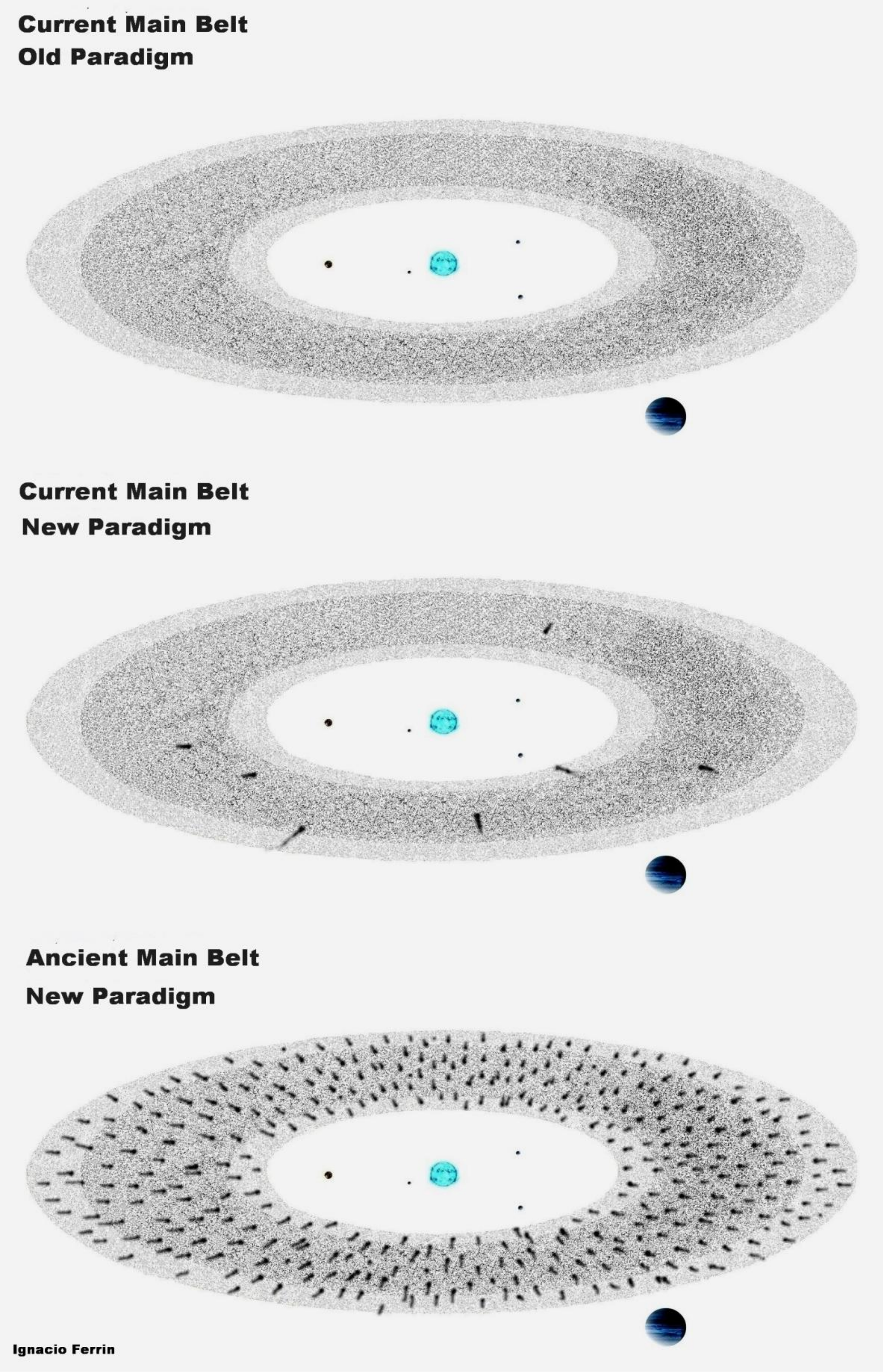

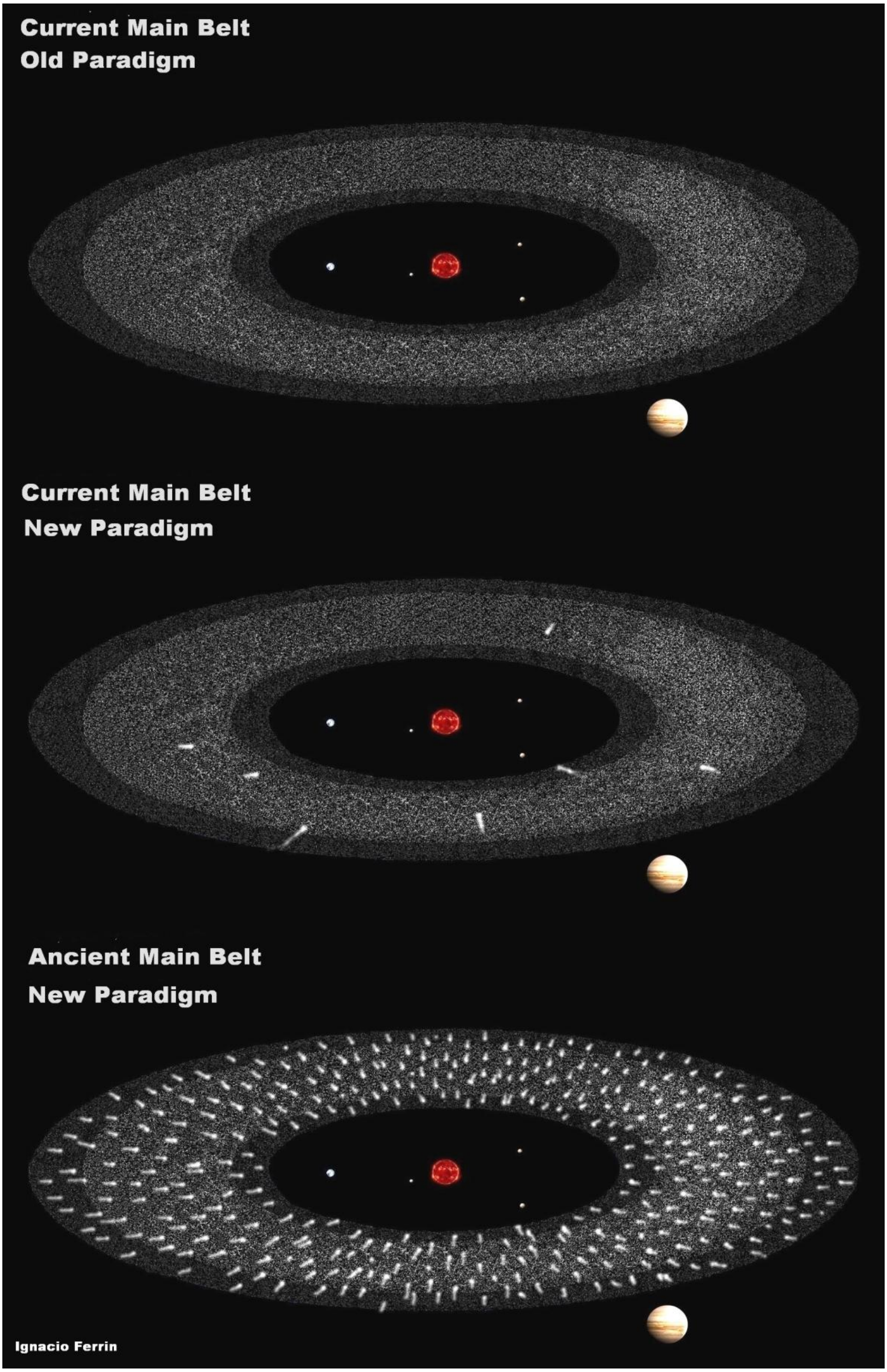